\documentclass{article}

\usepackage{qcircuit}
\usepackage{amssymb}
\usepackage{amsmath}
\PassOptionsToPackage{hyphens}{url}
\usepackage[breaklinks=true,colorlinks=true,linkcolor=blue,urlcolor=blue,citecolor=blue]{hyperref}
\usepackage[square,numbers,sort&compress]{natbib}
\usepackage{doi}
\usepackage{enumitem}
\usepackage{caption}
\usepackage[pagewise]{lineno}
\usepackage[nottoc]{tocbibind}
\usepackage{perpage}
\MakePerPage{footnote}

\usepackage{tikz}
\usepackage{tikzscale}
\usetikzlibrary{positioning}
\usetikzlibrary{calc}
\usetikzlibrary{decorations.pathreplacing,angles,quotes}

\usepackage{algorithm}
\usepackage{algpseudocode}
\floatname{algorithm}{Protocol}

\usepackage{pgfplots}
\usepackage{pgfplotstable}
\pgfplotsset{compat=1.13}

\usepackage{csquotes}
\usepackage{authblk}

\usepackage[a4paper, total={5.5in, 8in}]{geometry}

\newcommand{\ket}[1]{\left| #1 \right\rangle}
\newcommand{\bra}[1]{\left\langle #1 \right|}
\newcommand{\curlbrac}[1]{\left\{ #1 \right\}}

\newcommand{\figref}[1]{Figure \ref{#1}}
\newcommand{\defref}[1]{Definition \ref{#1}}
\renewcommand{\eqref}[1]{equation (\ref{#1})}
\newcommand{\secref}[1]{Section \ref{#1}}
\newcommand{\apref}[1]{\ref{#1}}
\newcommand{\theref}[1]{Theorem \ref{the:#1}}
\renewcommand{\algref}[1]{Protocol \ref{alg:#1}}

\newcommand{\breakalg}	{\vspace{2mm}

				\it Protocol continues below...

					\algstore{temp}
				\end{algorithmic}
			\end{algorithm}
			
			\begin{algorithm}
				\ContinuedFloat
				\caption{Continued}
				\begin{algorithmic}
					\algrestore{temp}
			}

\newcommand{\myinclude}[3]{\includegraphics[width=#1\linewidth, height=#2\linewidth]{#3}}

\newcommand{\myitem}[2]{\item[#1 - \emph{#2}:\label{itm:#1}]}

\newcommand{\IQP}{{\sf IQP}}
\newcommand{\BQP}{{\sf BQP}}

\newcommand{\PH}{{\sf PH}}

\newtheorem{theorem}{Theorem}[section]

\newtheorem{definition}{Definition}[section]

\newcommand{\lone}{$\ell_1$-norm distance}

\title{Methods for Classically Simulating Noisy Networked Quantum Architectures}

\author[1,2]{Iskren Vankov}
\affil[1]{School of Informatics, University of Edinburgh, 10 Crichton Street, Edinburgh EH8 9AB, UK}
\affil[2]{Department of Computer Science, University of Oxford, Parks Road, Oxford OX1 3PH, UK}
\author[1]{\thanks{Corresponding author: \href{mailto:daniel.mills@ed.ac.uk}{daniel.mills@ed.ac.uk}}Daniel Mills }
\author[1]{Petros Wallden}
\author[1,3]{Elham Kashefi}
\affil[3]{ Laboratoire d'Informatique de Paris 6, CNRS, UPMC - Sorbonne Universit\'es, 4 place Jussieu, 75005 Paris}
\begin{document}

\maketitle

\begin{abstract}

As research on building scalable quantum computers advances, it is important to be able to certify their correctness. Due to the exponential hardness of classically simulating quantum computation, straight-forward verification through classical simulation fails. However, we \emph{can} classically simulate small scale quantum computations and hence we \emph{are} able to test that devices behave as expected in this domain. This constitutes the first step towards obtaining confidence in the anticipated quantum-advantage when we extend to scales that can no longer be simulated. 

Realistic devices have restrictions due to their architecture and limitations due to physical imperfections and noise. Here we extend the usual ideal simulations by considering those effects.  We provide a general methodology for constructing realistic simulations emulating the physical system which will both provide a benchmark for realistic devices, and guide experimental research in the quest for quantum-advantage.

We exemplify our methodology by simulating a networked architecture and corresponding noise-model; in particular that of the device developed in the Networked Quantum Information Technologies Hub (NQIT) \cite{NQIT,Beaudrap}. For our simulations we use, with suitable modification, the classical simulator of \cite{Bravyi2016}. The specific problems considered belong to the class of Instantaneous Quantum Polynomial-time (\IQP{}) problems \cite{Shepherd2009}, a class believed to be hard for classical computing devices, and to be a promising candidate for the first demonstration of quantum-advantage. We first consider a subclass of \IQP{}, defined in \cite{Bermejo-Vega2017}, involving two-dimensional dynamical quantum simulators, before moving to more general instances of \IQP{}, but which are still restricted to the architecture of NQIT.

\end{abstract}

\noindent{\it Keywords}: \IQP{}, Classical simulation, NQIT, quantum-advantage, noise

\tableofcontents
\newpage

\section{Introduction}
\label{sec:intro}

Arguably the most significant developments in quantum technology would be that of devices for universal quantum computation, quantum simulation, or more bespoke tasks. These are likely to be disruptive innovations as they can, theoretically, provide an exponential speed-up in solving certain problems, as well as smaller advantages in other areas \cite{montanaro2016quantum}. The first implementation of these protocols will likely be quite some time in the future. Before then a first important milestone is to provide examples and proof-of-principle demonstrations of \emph{some} advantage being achieved with existing technologies \cite{preskill2018quantum, Arute2019}. This area of research has been termed the \emph{quantum-advantage problem}\footnote{This problem has come to be known by many names which are regularly used interchangeably. In particular ``quantum-supremacy'' or ``quantum-superiority'' are also popular. See \cite{pontiff, zeng} for some discussion on the matter.} \cite{harrow2017quantum, preskill2012quantum}. 

The goal when demonstrating quantum-advantage is to \emph{prove} one gains an advantage in solving a set of problems by using a device which utilises quantum mechanics. This advantage is measured relative to solving the same set of problems using \emph{any} available purely classical machine, implicitly ensuring that a device with this property utilises some quantum capabilities. Given a device, one may therefore say that it has demonstrated quantum-advantage by disproving the following hypothesis.

\begin{center}
	\it For any problem, there is a classical machine performing as well or better at solving the problem than the given device.
\end{center}

Providing a means of certifying an advantage has been achieved is of the utmost importance. This is a sub case of the more general problem of verifying a quantum computation has been implemented as expected \cite{gheorghiu2017verification, fitzsimons2017private}. The general problem is solved if one allows either for the verifier\footnote{In this work we will refer to the \emph{verifier} as the entity verifying the \emph{prover}.} to have a small quantum computer \cite{broadbent2015verify, fitzsimons2012unconditionally, fitzsimons2015post, aharonov2017interactive}, the verifier to interact classically with two non-communicating quantum devices (either both universal quantum provers \cite{reichardt2012classical} or one a universal quantum computer and one a measuring device \cite{GKW15}), or for computational complexity conjectures to be made \cite{mahadev2018classical, qfactory}. While these are remarkable results, they require either, or both, many more qubits than are required by an unverified implementation of a computation, or for the verifier to have quantum capabilities and a quantum communication channel to and from the prover. These requirements are not met by many of the most promising current technologies attempting to demonstrate a quantum-advantage.

In the absence of the general verification schemes, classical simulation can be invaluable. While it cannot reproduce large quantum computations, the technique can reproduce the behaviour of small instances. We can then compare these simulations with experimental results to confirm that the behaviour matches our predictions. By scaling our simulations beyond what is experimentally possible, and towards the regime of quantum-advantage, we can predict and prepare for the device's behaviour in this domain and understand how near term devices perform when implementing quantum-advantage protocols \cite{boixo2016characterizing}. Indeed by pushing our simulations to their limit we understand what is classically possible, giving a lower bound on the scale at which we would expect to observe a quantum-advantage for a given computation \cite{neville2017no, pednault2017breaking, Clifford:2018:CCB:3174304.3175276, chen2018classical, guerreschi2018qaoa, markov2018quantum, Villalonga2019Establishing, pednault2019leveraging}. Ultimately, we hope to address the following question.

\begin{center}
	\it Can we expect to observe a quantum-advantage using a given computation and architecture?
\end{center}

In quantum mechanics `the total is greater than the sum of it's parts' so testing small components of a quantum system is not sufficient to make precise predictions about it's behaviour at larger scales. This applies to testing small problem instances too. That being said, by simulating systems of size as close as possible to the classical limit we may more assuredly extrapolate that the device functions as modelled in the quantum-advantage regime. This is firstly because by testing, for example, modules 20-qubits in size, we are more confident that the phenomena we identify will manifest in larger systems than if we had tested single or two qubit modules. Secondly, since the regime of quantum-advantage is by definition just beyond the realm of classical simulation it is reasonable to assume phenomena in the realm where classical simulation is possible, but close to the quantum-advantage realm, exist in some form in the quantum-advantage domain.

For us to make reasonable predictions, our simulations must mimic the limitations of physical implementations. Arguably, chief among these limitations is noise. Currently the cost of fault tolerance \cite{Preskill385, RevModPhys.87.307, campbell2017roads} is high so early demonstrations of quantum-advantage will likely involve imperfect logical qubits\footnote{\emph{Logical qubits}, which are manipulated by the protocol, are often built of several \emph{physical qubits} using error correction codes.}. The quantum-advantage problem then becomes more subtle as the noise could destroy the advantage expected in a perfect run. A trade-off between the cost of removing the noise and, if it is not removed, the possible diminishing of the advantage of using a quantum computer, must be evaluated. 

Here we explore the impact of noise on the shape of distributions produced by quantum computers, but not if the noisy distributions are hard to reproduce classically. Indeed, in some cases it is known a small amount of noise can destroy the quantum-advantage \cite{Bremner2017achievingsupremacy}. Even in this case, classical simulation can be valuable. By varying noise levels in the simulations we can determine which types of imperfections lead to the greatest deviation from the perfect output. We can then suggest experimental groups prioritise improvements on those imperfections. 

While we find classical simulation to be an invaluable tool, other complementary benchmarking techniques have been explored. For example, randomised benchmarking \cite{PhysRevA.77.012307} and benchmarking using specific applications \cite{PhysRevA.99.032306, McCaskey2019Chemistry, cervera2019quantum} utilise the quantum technology directly. Other supremacy focused metrics utilising bespoke protocols \cite{aaronson2016complexity, boixo2016characterizing} have also been introduced, as well as metrics which consider a series of measures of error rate, qubit count, possible circuit depth etc to enable comparison between quantum devices \cite{Bishop2017QuantumV}. In line with the work of this paper, a recent pre-print utilises classical simulation to explore a selection of these methods by using the Summit supercomputer, the most powerful supercomputer at the time of writing \cite{Villalonga2019Establishing}.

In this work we give a methodology to follow when using classical simulation to benchmark quantum devices against the performance required for a demonstration of quantum-advantage and when guiding experiments pursuing a demonstration of quantum-advantage. We exemplify the methodology by considering \IQP{} problems implemented on the NQIT quantum device and show that the current size and noise-levels of the NQIT device make a demonstration of quantum-advantage unlikely. We further show that dephasing errors are the main source of degradation so recommend experimental labs prioritise reducing this type of error. We suggest, and simulate, an error-correction code, which corrects for these errors. Our results indicate that this approach improves performance considerably and makes a demonstration of quantum-advantage by implementing \IQP{} instances on the NQIT device more likely.

\secref{sec:method} contains the aforementioned methodology, which is then illustrated with examples in the following sections. In particular, in \secref{sec:preliminaries} we illustrate the technique, discussed in \secref{sec:roadmap}, for choosing the problems, architecture and simulator for our purposes. In \secref{sec:results} we illustrate the principles for numerical experiment design presented in \secref{sec:experiment_method} and: present simulations which can be used to benchmark the NQIT device, vary the noise levels in order to identify the main sources of error, and suggest steps to reduce these errors. We conclude in \secref{sec:discussion}.

\section{Methodology}
\label{sec:method}

Here we detail the methodology followed, addressing two areas. First, in \secref{sec:roadmap}, we give principles to follow when choosing a computational problem, experimental system, and classical simulator for the purpose of exploring quantum-advantage in near term devices. Second, in \secref{sec:experiment_method}, we give a methodology for designing numerical experiments, specifically when trying to assess the plausibility of a quantum-advantage demonstration. The methodology we introduce is sufficiently general as to be followed by other similar but original works. We will keep two desired outcomes in mind:

\begin{description}
	
	\myitem{Outcome 1}{Benchmark Device} By choosing parameters such as noise and problem size to be comparable with an actual experiment, we \emph{use the simulation to certify} the experiment/device and to predict it's performance as a demonstrator of quantum-advantage.
	
	\myitem{Outcome 2}{Feedback to Experimentalists} By altering the parameters we determine which imperfections have the greatest negative impact and \emph{provide advice} about which are the most urgent and beneficial hardware improvements.
	
\end{description}

\subsection{Problem, Architecture and Simulator Selection}
\label{sec:roadmap}

\begin{figure}
    	\centering
    	\begin{tikzpicture}[
    	    prob/.style={rectangle, draw=red!60, fill=red!5, very thick, inner sep=0.2cm, rounded corners = 4pt},
    	    input/.style={rectangle, draw=green!60, fill=green!5, very thick, inner sep=0.2cm, rounded corners = 4pt},
    	    output/.style={rectangle, draw=blue!60, fill=blue!5, very thick, inner sep=0.2cm, rounded corners = 4pt},
    	]
    		%Nodes
    		\node[prob] (problem) {Hard Problem};
    		\node[prob] (system) [below = of problem] {Experimental Setup};
    		\node[input] (gate) [right = of system] {Gate Set}; 
    		\node[input] (architecture) [left = of system] {Architecture};
    		\node[prob] (model) [below = of system] {Abstract Noise Model};
    		\node[input] (noise) [right = of model] {Noise}; 
    		\node[input] (time) [left = of model] {Gate Times}; 
    		\node[prob] (simulator) [below = of model] {Classical Simulator};
    		\node[output, align=center] (bench) [below right = of simulator] {Identify\\Improvements}; 
    		\node[output] (guide) [below left = of simulator] {Benchmark Simulator};
    		\node[output] (benchsim) [below = of simulator] {Benchmark Device};

    		%Lines
    		\draw[->, thick] (problem.south) -- (system.north);
    		\draw[->, thick] (system.south) -- (model.north);
    		\draw[<-, thick] (system.east) -- (gate.west);
    		\draw[<-, thick] (system.west) -- (architecture.east);
    		\draw[->, thick] (model.south) -- (simulator.north);
    		\draw[<-, thick] (model.east) -- (noise.west);
    		\draw[<-, thick] (model.west) -- (time.east);
    		
    		\draw[->, thick] (simulator.east) .. controls +(right:7mm) and +(up:7mm) .. (bench.north);
    		\draw[->, thick] (simulator.west) .. controls +(left:7mm) and +(up:7mm) .. (guide.north);
    		\draw[->, thick] (simulator.south) -- (benchsim.north);
    		
    		\draw[->, thick, dotted] (simulator.west) .. controls +(left:7mm) and +(left:7mm) .. (model.west);
    		\draw[->, thick, dotted] (model.east) .. controls +(right:7mm) and +(right:7mm) .. (system.east);
    		\draw[->, thick, dotted] (system.west) .. controls +(left:7mm) and +(left:7mm) .. (problem.west);
    	\end{tikzpicture}
    	\caption{The methodology proposed in this paper. The consideration of each step is preceded by its ancestor in the diagram, with feedback (dotted arrows) between steps, and contributing factors indicated from the sides. Outcomes are detailed at the base of the figure.}
    	\label{fig:method}
\end{figure}
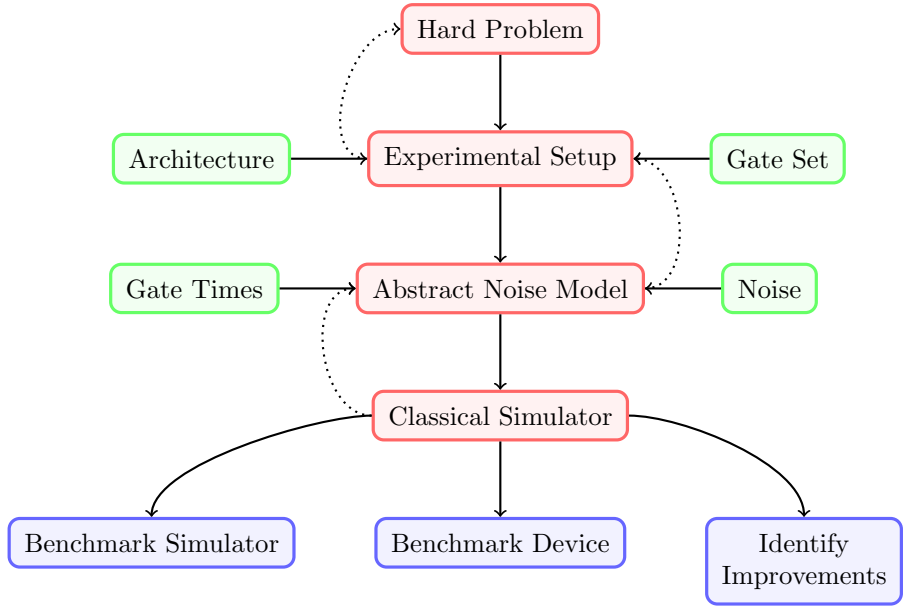

Here we give the method utilised in selecting the problem, experimental setup, and classical simulator used in achieving the above outcomes. We represent this methodology schematically in \figref{fig:method}.
\begin{description}

	\myitem{Step 1}{Hard Problem} Select a set of problems which: we know, or conjecture, to be classically hard; despite their hardness, need not be \BQP{}-complete (i.e. do not exhibit the full power of quantum computation) and are easier to implement than a universal quantum device; and show indications of the advantage in the quantum case persisting in the presence of noise.
	
\end{description}
It is reasonable to assume that the problem which first demonstrates quantum-advantage will fit the above description.
\begin{description}

	\myitem{Step 2}{Experimental Setup} Select an experimental set-up for which there exists reason to believing it could be built in the near term. Examine architecture restrictions including the quantum computation model (circuit, adiabatic, measurement-based, etc), the connectivity of the qubits, and the operations which are natural to the setting. 
	
	\myitem{Step 3}{Abstract Noise Model} Decide on a noise model to use, which should depend on the experimental implementation studied\footnote{For example the noise model for ion trap, photonic and superconducter implementations will be quite different} and on experimental measurements of the noise. For the quantum computation being considered, translate the noise into abstract operations appropriate for simulation. 

	\myitem{Step 4}{Classical Simulator} Select a classical simulator that is best suited for the problem under consideration. This is not, in general, a brute-force simulation and the specific choice can be such that it performs better for the problem, or instances there of, being considered.

\end{description}
While we consider each step in turn, we encourage feed-back between them. From the conclusions drawn at each step we ``tailor-make'' the construction of others.

\subsection{Numerical Experiment Design}
\label{sec:experiment_method}

Our analysis consists of three parts for each numerical experiment. In the first we test the suitability of the classical simulator we plan to use, while in the second we use the simulator and take into account realistic or projected noise. While the first part benchmarks the simulator, the second allows us to achieve Outcome 1 listed in the introduction to this section. The third part of the experiment involves altering the parameters to achieve Outcome 2.

\begin{description}
	\myitem{Part 1}{Simulator Benchmarking} Typically, the best classical simulators are probabilistic with errors which scale with the size of the computation. Therefore one must test the simulator chosen works as expected, specifically for the problem considered. Do this by running smaller instances of the problem and comparing the resulting distributions to a less efficient brute-force simulation. In particular: 
	\begin{itemize}
		\item Generate random small instances of the problem.\footnote{Here the problem that we simulate need not be hard as we are simply benchmarking the simulator, and not the prospect for quantum-advantage. The hard problem we consider should, however, be a subset of the general class we simulate here.}
		\item Complete a brute-force simulation of the generated problem.
		\item Adapt our chosen simulator to solve those instances, and solve many times.
		\item Compare the brute-force and aggregated simulator outcomes.\footnote{The simulator we use has a non-deterministic outcome so we take the average or `aggregated simulator outcome' as a means to compare.}
	\end{itemize}
    In this way we establish the simulator's accuracy.
    
	\myitem{Part 2}{Device Benchmarking} To address Outcome 1, impose constraints reflecting the implementation. Where possible, compare these simulations with experiments to determine the accuracy of any predictions made. Use the following steps:
	\begin{itemize}
		\item Generate random instances of the problem, restricted to the architecture.
		\item Generate many random instances of noise to generate many noisy circuits.
		\item Solve each noisy circuit and the original perfect circuit many times.
		\item Compare the aggregated simulations in the perfect case and the average of the aggregated noisy simulations.
		\item Use suitable parameters and compare with actual experimental realisations.
	\end{itemize} 
	In this way one can estimate the noise's influence.	
	
	\myitem{Part 3}{Guiding Future Experiments} Impose constraints coming from the realistic setting to the simulation and compare results with exploratory simulations with varying noise levels. This comparison is done to obtain an indication of the speed at which the noise ``corrupts'' the computation. Use this as a tool to provide feedback to experimental groups about which aspects of their devices they should prioritise improving. In so doing, we address Outcome 2.	
	\begin{itemize}
		\item Proceed as in Part 2 but with a varied noise model.
		\item Compare these results with simulations using the original noise model to understands the impact of the new noise model.
		\item If some change to the noise model is shown to result in a large improvement of the quality of the computation:
		\begin{enumerate}
		    \item Feed this information back to experimentalists so that they can prioritise reducing this type of noise
		    \item Consider theoretical methods to mitigate this specific type of error and test the performance in simulations. For example, introducing partial error-correction to deal with the single most important source of error. 
		\end{enumerate}
	\end{itemize}
\end{description}
While each part builds on from its predecessor, and so should follow it in the order of experiments, we may stop at some part if proceeding would not be advantageous.

We will not compare our results to those of experimentalists, as we describe above. However we recognise this as an important step and hope to do so in future work. Here we focus on using classical simulation to make predictions about the impact of noise.

\section{Exemplifying the Problem, Architecture, and Simulator Selection Methodology}
\label{sec:preliminaries}

Following the methodology for selecting a problem, architecture and simulator, discussed in \secref{sec:roadmap}: in \secref{sec:IQPintro} we present the class of problems considered; in \secref{sec:NQIT_arc} and \secref{sec:noise}, the physical system investigated; and in \secref{sec:class_sim}, the classical simulation technique used.

\subsection{Step 1 : The Instantaneous Quantum Polytime Machine}
\label{sec:IQPintro}

Step 1 of \secref{sec:roadmap} concerns the problems to consider during our simulations, which in our case will belong to the \IQP{} (Instantaneous Quantum Polynomial time) class \cite{Shepherd2009,Shepherd2010}. \IQP{} is a non-universal class of quantum computations which, like the one clean qubit model \cite{knill1998power,morimae2014hardness}, the boson sampling model \cite{aaronson2011computational, gard2015introduction, aaronson2013bosonsampling}, the Ising model \cite{gao2017quantum,van2008completeness}, etc, is thought to be able to demonstrate quantum-advantage, while also being designed with the goal of early implementation in mind. Indeed, current predictions \cite{howmanyqubits} put the number of qubits one expects to require for a demonstration of quantum-advantage using \IQP{} within the realm of what is thought to be possible in the near future.

\IQP{} circuits consist of commuting gates, a property which could theoretically be used to parallelise the computation and reduce the, physically hard to achieve, requirement for quantum memory\footnote{Quantum memory is hard to achieve in the sense that it is hard to store quantum states for long periods of time without them succumbing to noise.}. As well as being easier to implement, \IQP{} is believed to be hard to simulate classically \cite{bremner2010classical}, even in some relaxed settings. It remains hard in the approximate case \cite{Bremner2015}, when one imposes extra restrictions on the circuits \cite{Hoban2014}, or even in the presence of noise \cite{Bremner2017achievingsupremacy, fujii2014computational}. There also exists efficient methods for verifying some \IQP{} computations without classical simulation \cite{Shepherd2009, mills2017information, hangleiter2017direct}.

The existence of a possible demonstration of quantum-advantage under these very restrictive settings makes the \IQP{} class an exciting one to explore. We defined the class formally in \secref{sec:IQPdefinition}, explore related hardness results in \secref{sec:iqp_thms}, derive a concrete implementation in \secref{sec:IQP in MBQC} and explore an example of a particular set of problems that meet the hard problem selection conditions of Step 1 of \secref{sec:roadmap} in \secref{sec:2d-dqs}.

\subsubsection{Formal Definitions}
\label{sec:IQPdefinition}

An \IQP{} machine is defined by its capacity to implement $X$-programs and sample from the output distribution.
\begin{definition}[$X$-program]
	An \emph{$X$-program} consists of a Hamiltonian comprised of a sum of products of $X$ operators on different qubits, and $\theta\in[0,2\pi]$ describing the time for which it is applied.  The $h^{\mathrm{th}}$ term of the sum has a corresponding vector $\mathbf{q}_{h} \in \left\{ 0 , 1 \right\} ^{n_{a}}$, called a \emph{program element}, which defines on which of the $n_a$ input qubits, the product of $X$ operators which constitute that term, acts. The vector $\mathbf{q}_{h}$ has 1 in the $j^{\mathrm{th}}$ position when $X$ is applied on the $j^{\mathrm{th}}$ qubit. As such, we can describe the $X$-program using $\theta$ and the matrix $\mathbf{Q}=(\mathbf{Q}_{hj})\in\{0,1\}^{n_g\times n_a}$ which has as rows the program elements $\mathbf{q}_{h}$, $h=1,\dots,n_g$.
\end{definition}

Applying the $X$-program defined above to the state $\ket{0^{n_a}}$ and measuring the result in the $Z$ basis produces the following probability distribution, $\mathtt{X}$, of outcomes:
\begin{equation}
	\label{equ:IQP probability distribution}
	\mathbb{P} \left( \mathtt{X} = \widetilde{x} \right) = \left| \bra{\widetilde{x}} \exp \left( \sum_{h=1}^{n_g} i \theta \bigotimes_{j: \mathbf{Q}_{hj} = 1} X_{j}\right) \ket{0^{n_{a}}}  \right|^2 , \quad \widetilde{x} \in \{0,1\}^{n_{a}}
\end{equation}

\begin{definition}[The \IQP{} machine]
	\label{def:IQP oracle}
	Given an $X$-program, an \emph{\IQP{} machine} is any computational method capable of efficiently returning a sample $\widetilde{x}$ from the probability distribution of \eqref{equ:IQP probability distribution}.
\end{definition}

\subsubsection{Hardness Results and Their Robustness to Noise}
\label{sec:iqp_thms}

\IQP{} involves only gates which are diagonal in the Pauli-$X$ basis so does not achieve the full power of quantum computation. However, it is believed to be hard to classically simulate. Below we consider \emph{weak simulation} of a circuit family which is when, given a circuit's description, its output distribution can be sampled from by a polynomial time classical computer.
\begin{theorem}[informal from \cite{bremner2010classical}]
	\label{the:IQP mult hardness}
    If the output probability distributions generated by uniform families of \IQP{} circuits could be weakly classically simulated then the polynomial hierarchy (\PH{}) \cite{Stockmeyer1976} would collapse to its third level.
\end{theorem}
A collapse of \PH{} is thought to be unlikely, giving us confidence in the hardness of \IQP{}.

While \theref{IQP mult hardness} is a worst case hardness result, we can trim some instances form the set of problems we would expect to demonstrate quantum-advantage. For example \cite{Shepherd2009} when $\theta \in \left\{ \frac{\pi n}{4} : n \in \mathbb{Z} \right\}$ the result of the computation is classically computable. In the protocols we set $\theta = \frac{\pi}{8}$, giving us the necessary hardness.

\theref{IQP mult hardness} and similar results in \cite{Hoban2014} are remarkable in their demonstration that quantum computers which are very much weaker than a universal \BQP{} machine are impossible to classically simulate. These results are, however, proven in the setting where one demands a classical simulator produce samples which are within a multiplicative error, which depends on the probability of the sample, of the ideal quantum distribution. It is more realistic, and closer to the true capabilities of noisy quantum computers, to allow the classical simulator to be wrong up to an additive error. That is to say that the device need not necessarily sample from the ideal distribution $P$, but any distribution $\widetilde{P}$ with the property 
\begin{equation*}
    \sum_{x \in \left\{ 0 , 1 \right\}^{n_a}} \left| \widetilde{P} \left( x \right) - P \left( x \right)\right| \leq \epsilon
\end{equation*}
for some constant $\epsilon$. This measure of distance between distributions is also called the \lone. In this case too, hardness results exists.

\begin{theorem}[informal from \cite{Bremner2015}]
	Assume either one of two conjectures, relating to the hardness of Ising partition function and the gap of degree 3 polynomials, and the stability of the \PH{}, it is impossible to classically sample from the output probability distribution of any \IQP{} circuit in polynomial time, up to an additive error of $\epsilon = \frac{1}{192}$.
\end{theorem}

We will take the hardness of weak simulation up to additive error as an indication that a class of problems is promising for an early demonstration of quantum-advantage. This is justified because it seems plausible that noise will have a similar impact on average case problems, which we simulate, and worst case problems, for which hardness results exist. Thus we can draw conclusions about the impact of noise on the hard cases from its impact on average cases.

Ideally, we would like for our class to demonstrate an advantage in the average case as proofs of these results are often constructive, and would present us with schemes to implement. Such results, of which the following is an example, are harder to obtain, especially if one requires noise tolerance and architectural restrictions.

\begin{theorem}[informal from \cite{Bremner2017achievingsupremacy}]
    \label{the:2d iqp hard}
    Assuming the integrity of \PH{} and the difficulty of approximating an Ising model partition function; there is a family of \IQP{} circuits, implemented in depth $O ( \sqrt{n} \log n )$ on a 2D square lattice and containing $O ( n \log n )$ 2-qubit gates, for which a constant fraction of circuits cannot be simulated classically.
\end{theorem}

Here the simulation is understood as a simulation up to an additive error. This 2D square lattice architecture is favoured by many quantum computers today \cite{NQITreport2016, sete2016functional} and while we hope to be impartial to the architecture \cite{beals2013efficient, RLrouting, cowtan2019qubit}, for early devices it is important to engineer our tests with this in mind. However, it is likely that the qubit routing used to implement the circuit of \theref{2d iqp hard} on a square lattice requires many swap gates. These would not commute with the rest of the circuit, destroying the instantaneous nature which, as we will see, we prefer for our purposes.

Theoretical studies of quantum-advantage in the presence of noise have also explored the following, arguably more realistic, settings. The first considers independent depolarising noise which is added to all qubits at the end of the circuit. In this case the noise per qubit does not, as in the additive case, depend on the number of qubits. It is shown \cite{Bremner2017achievingsupremacy} that the circuit family of \theref{2d iqp hard} are classically simulable in this noise model but that classical hardness can be recovered by modifying the circuits to include some classical error correction technique. Second, the more general case of \emph{independent} noise being applied to each gate also leads to a wide family of circuits becoming classically simulable \cite{gao2018efficient}.

In our work, we do not explore the impact of noise on the quantum-advantage at a theoretical level, as was done in the aforementioned works, but suggest that numerical exploration should be done in parallel with the theoretical analysis. This would guide us in understanding which realistic experimental setting is best to demonstrate quantum-advantage with \IQP{} problems. 

%%%%%%%%%%%%%%%%%%%%%%%%%%%%%%%%%%%%%%%%%%%%%%%%%%%%%%%%%%%%%%%%%%%%%%%%%%%%%%%%%%%

\subsubsection{IQP-MBQC: A Measurement Based Implementation}
\label{sec:IQP in MBQC}

A common framework for studying quantum computation is the Measurement-Based Quantum Computation (MBQC) model \cite{raussendorf2001one, raussendorf2003measurement, Raussendorf2005}. Problems in the \IQP{} class admits a realisation, using MBQC, which is particularly useful since it explicitly parallelises the computation. 

The MBQC implementation of a given $X$-program uses a graph state defined by a corresponding bipartite graph.

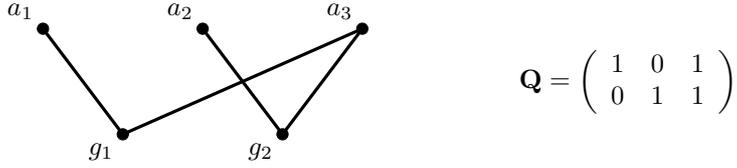
\begin{figure}[!ht]
	\centering
	\begin{tikzpicture}[scale = 0.7]
		\filldraw (1.5,0) circle (3pt) node[anchor = north east] {$g_{1}$};
		\filldraw (4.5,0) circle (3pt) node[anchor = north east] {$g_{2}$};
		
		\filldraw (0,2) circle (3pt) node[anchor = south east] {$a_{1}$};
		\draw[very thick] (0,2) -- (1.5,0);
		
		\filldraw (3,2) circle (3pt) node[anchor = south east] {$a_{2}$};
		\draw[very thick] (3,2) -- (4.5,0);
		
		\filldraw (6,2) circle (3pt) node[anchor = south east] {$a_{3}$};
		\draw[very thick] (6,2) -- (1.5,0);
		\draw[very thick] (6,2) -- (4.5,0);
		
		\node at (11,1) {$\mathbf{Q} =   \left( 
		                                    \begin{array}{ccc}
                                                1 & 0 & 1 \\
                                                0 & 1 & 1
                                            \end{array}
                                        \right)$};
	\end{tikzpicture}
	\caption{An example of an bipartite graph described by matrix $\mathbf{Q}$. Here, $n_{a} = 3$ and $n_{g} = 2$ while the partition used is $ A = \left[ a_1 , a_2 , a_3 \right]$ and $G = \left[ g_1 , g_2 \right]$.}
	\label{fig:bipartite graph}
\end{figure}

\begin{definition}[Bipartite graph]
	\label{def:IQP graph}
	We define the \emph{bipartite graph} of an $X$-program $(\mathbf{Q},\theta)$, as the graph with biadjacency matrix $\mathbf{Q}=(\mathbf{Q}_{hj})\in\{0,1\}^{n_g\times n_a}$. This means that there is a bipartition of vertices into two sets $A$ and $G$ of cardinality $n_a$ and $n_g$ and that an edge exists in the graph between vertex $g_h$ of set $G$ and vertex $a_j$ of set $A$ when $\mathbf{Q}_{hj}=1$. The sets of vertices $G = \left\{ g_{1} , ... , g_{n_{g}} \right\}$ and $A = \left\{ a_{1} , ... , a_{n_{a}} \right\}$ will be called \emph{gate} and \emph{application vertices} respectively. See Figure \ref{fig:bipartite graph} for an example.
\end{definition}

One can prove \cite{Shepherd2009} that the distribution of \eqref{equ:IQP probability distribution} can be achieved by initialising $n_a$ \emph{application qubits} in the states $\ket{a_{j}} = \ket{+}$, $n_g$ \emph{gate qubits} in the states $\ket{g_{h}} = \ket{+}$, applying Controlled-$Z$ operations between qubits when there is an edge in the bipartite graph described by the $X$-program matrix $\mathbf{Q}$ and measuring the resulting state. The measurement of the application qubits is in the Hadamard basis, and of the gate qubits is in the basis of \eqref{equ:IQP measuremnt basis}.
\begin{equation}
	\label{equ:IQP measuremnt basis}
	\left\{ \ket{0_{\theta}} , \ket{1_{\theta}} \right\}= \left\{ \frac{1}{\sqrt{2}} \left(  e^{-i\theta}\ket{+} + e^{i\theta}\ket{-} \right) ,\frac{1}{\sqrt{2}} \left(e^{-i\theta} \ket{+} - e^{i\theta}\ket{-} \right)  \right\} 
\end{equation}
The measurement bases do not depend on the outcomes of other measurements and therefore can be parallelised to one round of entanglement and measurement. 

Importantly the distribution of \eqref{equ:IQP probability distribution} is achieved via this implementation in polynomial time. As such the complexity results of \secref{sec:iqp_thms} apply here.

%%%%%%%%%%%%%%%%%%%%%%%%%%%%%%%%%%%%%%%%%%%%%%%%%%%%%%%%%%%%%%%%%%%%%%%%%%%%%%%%%%%%

\subsubsection{Using the 2D-DQS Protocol to Demonstrate Quantum-Advantage}
\label{sec:2d-dqs}

In \cite{Bermejo-Vega2017} a subclass of \IQP{} problems called 2-dimensional dynamical quantum simulators (2D-DQS) are defined. The name references the 2D square lattice architecture involved and that they could be realised with sub-universal quantum simulators. Architecture I from \cite{Bermejo-Vega2017} is seen in \algref{jens}.

\begin{algorithm}
\begin{algorithmic}[1]
	\State Choose $\tau \in \curlbrac{0 , 1}^{N_x \times N_y}$ uniformly at random.
	\State Initialise the product state:
	\begin{equation}
		\ket{\phi_{\tau}} = \bigotimes_{i=1}^{N = N_x \times N_y} \left( \ket{0} + e^{i \tau_{i} \frac{\pi}{4}} \ket{1} \right)
	\end{equation}
	\State Allow system to evolve for time $t = 1$ according to the nearest neighbour, translation invariant, Ising Hamiltonian:
	\begin{equation}
		H := \sum_{\left( i , j \right) \in E} \frac{\pi}{4} Z_{i} Z_{j} - \sum_{i \in V} \frac{\pi}{4} Z_{i}
	\end{equation}
	This is equivalent to applying controlled $Z$ operations on each edge.
	\State Measure all qubits in the $X$ basis.
\end{algorithmic}
\caption{A description of an instance of the 2D-DQS problem introduced by \cite{Bermejo-Vega2017}. $E$ and $V$ are the edge and vertex set respectively of a $N_x \times N_y$ 2D square lattice.}
\label{alg:jens}
\end{algorithm}

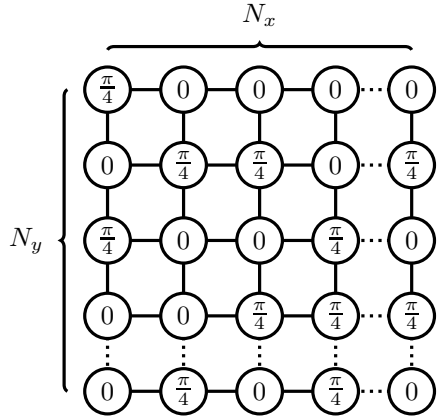
\begin{figure}%[ht]
	\centering
	\begin{tikzpicture}
		\draw[very thick] (0,1) -- (0,4);
		\draw[very thick] (1,1) -- (1,4);
		\draw[very thick] (2,1) -- (2,4);	
		\draw[very thick] (3,1) -- (3,4);
		\draw[very thick] (4,1) -- (4,4);
		\draw[very thick] (0,0) -- (3,0);
		\draw[very thick] (0,1) -- (3,1);
		\draw[very thick] (0,2) -- (3,2);
		\draw[very thick] (0,3) -- (3,3);
		\draw[very thick] (0,4) -- (3,4);
		
		\draw[very thick, dotted] (0,0) -- (0,1);
		\draw[very thick, dotted] (1,0) -- (1,1);
		\draw[very thick, dotted] (2,0) -- (2,1);	
		\draw[very thick, dotted] (3,0) -- (3,1);
		\draw[very thick, dotted] (4,0) -- (4,1);
		\draw[very thick, dotted] (3,0) -- (4,0);
		\draw[very thick, dotted] (3,1) -- (4,1);
		\draw[very thick, dotted] (3,2) -- (4,2);
		\draw[very thick, dotted] (3,3) -- (4,3);
		\draw[very thick, dotted] (3,4) -- (4,4);
	
		\filldraw[very thick, fill = white] (0,0) circle (0.3) node {0};
		\filldraw[very thick, fill = white] (0,1) circle (0.3) node {0};
		\filldraw[very thick, fill = white] (0,2) circle (0.3) node {$\frac{\pi}{4}$};
		\filldraw[very thick, fill = white] (0,3) circle (0.3) node {0};
		\filldraw[very thick, fill = white] (0,4) circle (0.3) node {$\frac{\pi}{4}$};
		\filldraw[very thick, fill = white] (1,0) circle (0.3) node {$\frac{\pi}{4}$};
		\filldraw[very thick, fill = white] (1,1) circle (0.3) node {0};
		\filldraw[very thick, fill = white] (1,2) circle (0.3) node {0};
		\filldraw[very thick, fill = white] (1,3) circle (0.3) node {$\frac{\pi}{4}$};
		\filldraw[very thick, fill = white] (1,4) circle (0.3) node {0};
		\filldraw[very thick, fill = white] (2,0) circle (0.3) node {0};
		\filldraw[very thick, fill = white] (2,1) circle (0.3) node {$\frac{\pi}{4}$};
		\filldraw[very thick, fill = white] (2,2) circle (0.3) node {0};
		\filldraw[very thick, fill = white] (2,3) circle (0.3) node {$\frac{\pi}{4}$};
		\filldraw[very thick, fill = white] (2,4) circle (0.3) node {0};
		\filldraw[very thick, fill = white] (3,0) circle (0.3) node {$\frac{\pi}{4}$};
		\filldraw[very thick, fill = white] (3,1) circle (0.3) node {$\frac{\pi}{4}$};
		\filldraw[very thick, fill = white] (3,2) circle (0.3) node {$\frac{\pi}{4}$};
		\filldraw[very thick, fill = white] (3,3) circle (0.3) node {0};
		\filldraw[very thick, fill = white] (3,4) circle (0.3) node {0};
		\filldraw[very thick, fill = white] (4,0) circle (0.3) node {0};
		\filldraw[very thick, fill = white] (4,1) circle (0.3) node {$\frac{\pi}{4}$};
		\filldraw[very thick, fill = white] (4,2) circle (0.3) node {0};
		\filldraw[very thick, fill = white] (4,3) circle (0.3) node {$\frac{\pi}{4}$};
		\filldraw[very thick, fill = white] (4,4) circle (0.3) node {0};

		\draw[decoration={brace,raise=15pt}, decorate, very thick] (0,4) -- node[above=20pt] {$N_x$} (4,4);
		\draw[decoration={brace,raise=15pt}, decorate, very thick] (0,0) -- node[left=20pt] {$N_y$} (0,4);
	\end{tikzpicture}
	\caption{An example of an instance of the 2D-DQS problem for quantum-advantage, detailed in \algref{jens} and introduced in \cite{Bermejo-Vega2017}. The value in each qubit describes the state of initialisation while the lines connecting them indicate the application of a controlled $Z$ gates between those qubits. Each qubit of the resulting state is measured in the Pauli $X$ basis.}
	\label{fig:archetecture I}
\end{figure}

The construction is summarised in \figref{fig:archetecture I}. One realises that this is within \IQP{} by noting either that it is simply a Bloch sphere rotation of the definition in \secref{sec:IQPdefinition} or that it is a constant depth commuting circuit on a 2D lattice. The related hardness result for this architecture is seen in \theref{2D-DQS hardness}.

\begin{theorem}[informal from \cite{Bermejo-Vega2017}]
	\label{the:2D-DQS hardness}
	Assuming three conjectures (one being the non-collapse of \PH), a classical computer cannot sample from the outcome distribution of the architecture of \algref{jens} up to an additive error of $\frac{1}{22}$ in time polynomial in $N_x,N_y$.
\end{theorem}

We note that this problem seems a good candidate for our purposes, as described in the hard problem selection methodology of Step 1 in \secref{sec:roadmap}, since it is hard to simulate classically \emph{and} is experimentally realisable in the near term. A further advantage of this scheme is that the authors of \cite{Bermejo-Vega2017} provide an explicit means for a client with a simple measurement device to verify the protocol. This is an important feature for extending the analysis beyond the limits were classical simulation is possible.

\subsection{Step 2 : NQIT Architecture}
\label{sec:NQIT_arc}

The second choice to make is the physical system that we consider (Step 2 of \secref{sec:roadmap}). We chose the Q20:20 device being developed by the Networked Quantum Information Technologies Hub NQIT \cite{NQIT}. In fact we will model this device as closely as possible so it will also determine our choice of the noise model, as discussed in Step 3 of \secref{sec:noise}.  

Networked architectures like NQIT, which combine matter degrees of freedom in modules which are entangled via photonic degrees of freedom, have two important advantages. Firstly, once the implementation of connections between modules is perfected, this architecture can easily scale without significant extra challenges. The second advantage is that this architecture can be combined easily with communication tasks. Many applications of quantum computation are likely to involve multiple parties, a setting to which networked architectures are best suited.

\begin{figure}%[ht]

	\centering

    	\begin{tikzpicture}[scale = 0.85, transform shape]
		
        	    \node at (0,0) (N1) {};
            	\node (N2) [above = 2 of N1] {};
            	\node (N3) [above = 2 of N2] {};
            	\node (N4) [right = 2 of N1] {};
            	\node (N5) [above = 2 of N4] {};
            	\node (N6) [above = 2 of N5] {};
            	\node (N7) [right = 2 of N4] {};
            	\node (N8) [above = 2 of N7] {};
            	\node (N9) [above = 2 of N8] {};

	    	    \node (N10) [right = 2 of N9] {};
		        \node (N11) [right = 1 of N10] {};
    	    	\node (N12) [right = 1 of N11] {};
    	    	\node (N13) [right = 1 of N12] {};
    	    	\node (N14) [right = 1 of N13] {};

    		    \node (N15) [right = 2 of N8] {};
    	    	\node (N16) [right = 1 of N15] {};
    	    	\node (N17) [right = 1 of N16] {};
    	    	\node (N18) [right = 1 of N17] {};
    	    	\node (N19) [right = 1 of N18] {};
            
            	\filldraw (N1) circle (3pt);
            	\filldraw (N2) circle (3pt);
            	\filldraw (N3) circle (3pt);
            	\filldraw (N4) circle (3pt);
            	\filldraw (N5) circle (3pt);
            	\filldraw (N6) circle (3pt);
            	\filldraw (N7) circle (3pt);
            	\filldraw (N8) circle (3pt);
            	\filldraw (N9) circle (3pt);
            	
            	\draw[very thick, ->] (N1) -- ($(N1)!0.4!(N2)$);
            	\draw[very thick, ->] (N1) -- ($(N1)!0.4!(N4)$);
            	\draw[very thick, ->] (N2) -- ($(N2)!0.4!(N1)$);
            	\draw[very thick, ->] (N2) -- ($(N2)!0.4!(N3)$);
            	\draw[very thick, ->] (N2) -- ($(N2)!0.4!(N5)$);
            	\draw[very thick, ->] (N3) -- ($(N3)!0.4!(N2)$);
            	\draw[very thick, ->] (N3) -- ($(N3)!0.4!(N6)$);
            	\draw[very thick, ->] (N4) -- ($(N4)!0.4!(N1)$);
            	\draw[very thick, ->] (N4) -- ($(N4)!0.4!(N5)$);
            	\draw[very thick, ->] (N4) -- ($(N4)!0.4!(N7)$);
            	\draw[very thick, ->] (N5) -- ($(N5)!0.4!(N4)$);
            	\draw[very thick, ->] (N5) -- ($(N5)!0.4!(N2)$);
            	\draw[very thick, ->] (N5) -- ($(N5)!0.4!(N6)$);
            	\draw[very thick, ->] (N5) -- ($(N5)!0.4!(N8)$);
            	\draw[very thick, ->] (N6) -- ($(N6)!0.4!(N3)$);
            	\draw[very thick, ->] (N6) -- ($(N6)!0.4!(N5)$);
            	\draw[very thick, ->] (N6) -- ($(N6)!0.4!(N9)$);
            	\draw[very thick, ->] (N7) -- ($(N7)!0.4!(N4)$);
            	\draw[very thick, ->] (N7) -- ($(N7)!0.4!(N8)$);
            	\draw[very thick, ->] (N8) -- ($(N8)!0.4!(N7)$);
            	\draw[very thick, ->] (N8) -- ($(N8)!0.4!(N5)$);
            	\draw[very thick, ->] (N8) -- ($(N8)!0.4!(N9)$);
            	\draw[very thick, ->] (N9) -- ($(N9)!0.4!(N8)$);
            	\draw[very thick, ->] (N9) -- ($(N9)!0.4!(N6)$);

		\draw[dotted, very thick] (N9) circle (6pt);
		%\draw[dotted, very thick] (N12) ellipse (3 and 1);

		\draw[dashed, very thick, rounded corners = 15pt] (6,5.25) rectangle (12.5,3.75);

		\draw[very thick] (N10) circle (4pt);
		\draw[very thick] (N11) circle (4pt);
		\draw[very thick] (N12) circle (4pt);
		\draw[very thick] (N13) circle (4pt);
		\draw[very thick] (N14) circle (4pt);

		\draw[very thick, dotted, <-] ($(N9)!0.6!(N10)$) -- ($(N9)!0.2!(N10)$);

		\draw[dotted, very thick] (N8) circle (6pt);
		%\draw[dotted, very thick] (N17) ellipse (3 and 1);

		\draw[dashed, very thick, rounded corners = 15pt] (6,3) rectangle (12.5,1.5);

		\draw[very thick] (N15) circle (4pt);
		\draw[very thick] (N16) circle (4pt);
		\draw[very thick] (N17) circle (4pt);
		\draw[very thick] (N18) circle (4pt);
		\draw[very thick] (N19) circle (4pt);

		\draw[very thick, dotted, <-] ($(N8)!0.6!(N15)$) -- ($(N8)!0.2!(N15)$);

		\draw[very thick, dotted] (N10) -- (N15);
		\draw[very thick, dotted] (N11) -- (N16);

		\draw[very thick] (N12) -- (N17);

		\draw[very thick] (N18) -- (N17);
		\draw[very thick] (N13) -- (N14);

		\draw[decoration={brace,raise=10pt}, decorate, very thick] (N3) -- node[above=15pt] {$N_x$} (N9);
		\draw[decoration={brace,raise=10pt}, decorate, very thick] (N1) -- node[left=15pt] {$N_y$} (N3);
		\draw[decoration={brace,raise=30pt}, decorate, very thick] (N10) -- node[above=40pt] {$K$} (N14);
		\draw[decoration={brace,raise=30pt}, decorate, very thick] (N19) -- node[below=40pt] {$K'$} (N17);
   	\end{tikzpicture}

	\caption{Architecture of the NQIT device. We see on the left the connectivity between ion traps, and on the right an expanded view of individual ion traps and their internal and external connectivity. Dotted lines between ions in different ion traps in the expanded view indicate lower fidelity entanglement which is used to distil a higher fidelity entanglement indicated by the solid line. Note that $N = N_x N_y$.}

	\label{fig:NQIT arc}
\end{figure}
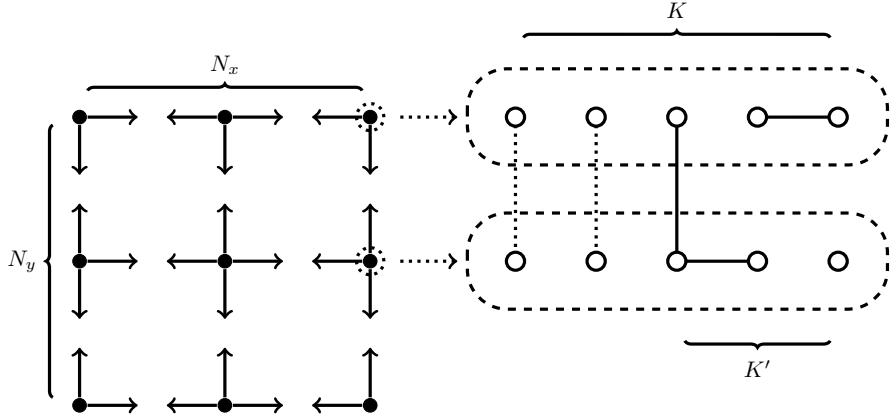

The device that NQIT is developing\footnote{Since the beginning of the project other variations of this architecture have been considered.} is called Q20:20. It consists of $N=20$ ion traps \cite{bruzewicz2019trapped} with $K=20$ ions (physical qubits) in each. Traps are arranged on a 2D grid with only nearest-neighbour interactions allowed, giving a maximum number of connections $D=4$. Different ion-traps are connected via high-fidelity entanglement between dedicated \emph{linking qubits}. This high-fidelity entanglement is realised through \emph{entanglement distillation} \cite{Bennett1996, Nigmatullin2016a} and consumes some of the physical qubits of each ion-trap, leaving $K'<K$ available qubits, before considering the cost of potential error-correction. Two-qubit gates between ion-traps can be applied by \emph{teleporting} the qubits into the same cell. Single and two-qubit gates within a single  ion-trap take place in special \emph{gate zones}. A summary of this information can be seen in \figref{fig:NQIT arc}.

These details are based on information obtained early in the NQIT project \cite{Beaudrap}. Since the project is still underway, the system parameters $N$, $K$, $K'$, $D$, and others, may change \cite{Nigmatullin2016a} and so we let them vary in our simulation toolbox.

Like the architecture itself, the operations that are possible on the NQIT device may vary. We select to use the following set:

\begin{description}
	\item[Preparation and measurement:] It is possible to prepare qubits in the Hadamard basis and measure qubits in the computational basis.
	\item[Single qubit operations:] The possible single qubit operations consist of the Hadamard gate and rotations by arbitrary angles, about arbitrary axes in the $X-Z$ plane. For practical reasons the axes will likely be restricted to integer multiples of fractions of $\pi$. Here we will choose $\frac{\pi}{4}$ giving us access to $T$ gates.
	\item[Two qubit operations:] Here the controlled $Z$ gate, $CZ$, is permitted.
	\item[Operations between traps:] It is possible to create a bell pair $\ket{\phi} = \frac{1}{\sqrt{2}} \left( \ket{01} + \ket{10} \right)$ between traps.
\end{description}

The gate set used by the NQIT device may change but this set is a plausible one. It will at least result in compilation to circuits with a comparable gate count and execution time to the final choice; both key factors in determining the effect of noise.

\subsection{Step 3 : NQIT Noise}
\label{sec:noise}

Following Step 3 of \secref{sec:roadmap}, we give a brief summary of all types of noise, the degree to which they impact computations in the case of NQIT, and how we will model them. We divide the noise into time-based, which we model as occurring randomly in time on each physical qubit independently, and operation-based, which we model as occurring when an operator is applied, and is only applied to the qubits on which the gate acts. The values listed below are acquired through measurements of the NQIT device \cite{Beaudrap}.

\paragraph{Time-based Noise}

\begin{description}
    \item[Depolarising]
	    Caused by scattering of amplitudes of the electron's wave-function between different energy levels of the ion. Modelled by a random single-qubit Pauli on each qubit at a rate of $\approx 9 \times 10^{-4} s^{-1}$.
    
    \item[Dephasing]
	    Entanglement reduction that destroys data not stored in the standard basis. Modelled by $Z$ gate on each qubit at a rate of $\approx (7.2 \pm 1.4) \times 10^{-3} s^{-1}$.    
\end{description}
To simulate these noise channels we need the execution times of different operations:
\begin{itemize}
	\item Preparation - $1 - 1.5$ms
	\item Measurement - $2 - 2.5$ms
	\item Single or two-qubit operation within a trap - $0.5$ms \footnote{This set of operations includes moving the qubits to the \textit{gate zone}.}
	\item Linking between traps - $1 - 2$s \footnote{This timing information is for the case of 10 distillation qubits.}
\end{itemize}

\paragraph{Operation-based Noise}

\begin{description}
    \item[Preparation]
    Error probability in preparing a state. Modelled by Pauli $X$ at rate of $\approx 2 \times 10^{-4}$.
    
    \item[Measurement]
    Similarly to preparation, measurement is also noisy. Rate of $\approx 5 \times 10^{-4}$ to measure incorrectly any qubit, which corresponds to an $X$ gate.
    
    \item[Single-qubit gates]
    Random Pauli operator applied in addition to the single-qubit gate with probability $\approx (1.5 \pm 0.45) \times 10^{-6}$.

    \item[Two-qubit gates]
    Modelled by independent single-qubit random Pauli errors on both qubits, each with probability $\approx (5.5 \pm 3.5) \times 10^{-4}$ and a further two-qubit error $Z \otimes Z$ with probability $\approx 6 \times 10^{-5}$.

    \item[Linking operations]
    Depending on the amount of entanglement distillation used \cite{Nigmatullin2016a}, this error varies since it is determined by the fidelity of the entanglement. If 10-qubits are used for distillation, then the effect is approximately the same as the regular (same ion-trap) two-qubit gate \cite{Beaudrap}. Moreover, using more qubits for distillation would not improve the computation since the same ion-trap qubit gates will still have higher errors. 
\end{description}
This noise description is specific to the NQIT Q20:20 device. However, the structure is general and other versions of the NQIT device or other quantum devices are likely to have similar ``specifications''. Therefore the toolbox developed should be adaptable to other quantum computation devices. The reader may refer to \apref{app:noise functions} for a systematic description of the noise.

\subsection{Step 4 : Clifford + T simulator of Bravyi and Gosset}
\label{sec:class_sim}

The last choice is to determine the classical simulator we use (Step 4). We use the improved Clifford + T simulator of \cite{Bravyi2016}, which we introduce here. As we will discuss, for the IQP-MBQC and 2D-DQS problems in \secref{sec:IQPintro}, this appears to be the most promising classical simulator. 

While it is thought that classical simulation of universal quantum computation comes at the cost of exponential complexity \cite{aaronson2011computational, bremner2010classical}, compared to naive brute-force simulations there exist more efficient ways to classically simulate quantum systems. These techniques extend the domain of applicability of classical simulations, and for specific problems, enables simulations even for large instances. For example, by employing tensor networks \cite{vidal2003efficient, fried2017qtorch} the simulation of low entanglement computation becomes accessible while low amounts of interference gives the same result \cite{stahlke2014quantum}. Using the positivity of the Wigner function  \cite{PhysRevLett.109.230503} or the quasi probability representation \cite{1367-2630-14-11-113011} one can also obtain more efficient classical simulations. Monte Carlo simulations \cite{magesan2013modeling, PhysRevA.91.022335, PhysRevA.89.022306, bennink2017unbiased, PhysRevLett.115.070501} have been developed to simulate noisy systems.

The Gottesman-Knill theorem \cite{gottesman1998heisenberg} states that a \emph{Clifford} circuit, built from the gate from the set $\left\{ S , H , CNOT \right\}$ acting on computational basis states and measurements in the computational basis, can be efficiently simulated on a classical computer. This result has since been greatly extended and improved \cite{aaronson2004improved, bartlett2002efficient, anders2006fast, yoder2012generalization, garcia2013quipu}. While the Clifford gate set is not universal even for classical computations \cite{aaronson2004improved}, adding just the $T$-gate to the set makes it universal for \emph{quantum} computation. In \cite{Bravyi2016}, a classical simulator for the Clifford + $T$ gate set, with run time exponential in the number of $T$-gates\footnote{The exact expression has $2^{\beta t}$, where $\beta < \frac{1}{2}$ and $t$ is the number of $T$-gates.} but polynomial in the number of qubits and Clifford gates, is developed. This allows efficient simulation of circuits with a logarithmic number of $T$-gates. Furthermore, because of the small exponent, it enables the classical simulation of larger instances than regular ``brute-force'' simulators. Hence by restricting the frequency of T-gates in the instances of the \IQP{} problem we consider, we can simulate even larger numbers of qubits/circuits than we would otherwise be able to. 

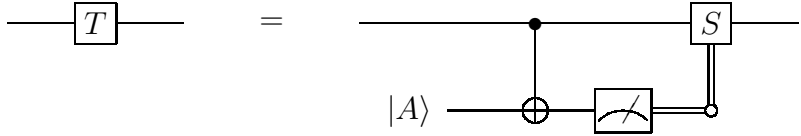
\begin{figure}
    \centering
    \large
    $
	\Qcircuit @C=1em @R=1em @! {
		& \gate{T} 	& \qw 	& = 	&  	& \qw 			& \ctrl{1} 	& \qw 		& \gate{S} 	& \qw \\
		& 		& 	&	&	& \lstick{\ket{A}}	& \targ		& \meter	& \controlo \cw \cwx	& 
	}
	$
    \caption{The gadget used to replace a T-gate \cite{zhou2000methodology}. $\ket{A} = \frac{1}{\sqrt{2}}(\ket{0} + e^{i\pi/4}\ket{1})$ are \textit{magic states} \cite{PhysRevA.71.022316}. }
    \label{fig:Tgadget}
\end{figure}

The details are given in \cite{Bravyi2016}, but here we give an outline of the idea. First all $T$ gates are replaced by the gadget of Figure \ref{fig:Tgadget}. The measurement is replaced by postselection onto the $0$ outcome and the magic state is replaced by a decomposition into exponentially many stabiliser states. These steps result in a purely stabiliser circuit and measurements of exponentially (in the $T$ count) many stabiliser states. The nature of the simulation means that Clifford gates can be simulated exactly, while the simulation of non-Clifford gates is probabilistic. Using this method the authors are able to simulate about $40$ qubits and $50$ $T$ gates in what they quote as `several hours'. Here we require the simulation of several thousand circuits and so we simulate fewer $T$ gates to allow this to be done in a reasonable time.

The instantiation of the simulator we use is a Clifford + $T$ gate set simulator from \cite{Bravyi2016} which produces the probability of measuring a \emph{single outcome}. In \cite{Bravyi2016} a more general simulator was also introduced which samples from the output distribution. The 2D-DQS problem chosen in \secref{sec:2d-dqs} is highly entangled, beyond stabiliser simulation and conveniently represented in the Clifford + $T$ gate set without the need for costly (in gate count) gate decomposition \cite{selinger2012efficient, PhysRevA.87.032332}\footnote{The work of \cite{Bravyi2016} has since been generalised \cite{GB2} to allow for other gate sets. Hence it may be possible for small instances of other protocols demonstrating quantum-advantage to be more efficiently simulated without being decomposed into the Clifford + $T$ gate set. For the scheme we use the Clifford + $T$ simulator is still optimal.}. This makes the simulator of \cite{Bravyi2016} perfect for our purposes, and others mentioned above less useful. There are many implementations of simulators available \cite{LaRose2018} but they are either more general purposes solutions \cite{smelyanskiy2016qhipster, steiger2016projectq, wecker2014liqui, cross2018ibm, smith2016practical, QSharp, jones2018QuEST, garcia2013quipu, villalonga2018flexible, Q++}, which can mean a large overhead for our specific set of circuits, or bespoke for tasks other than the one we require here \cite{mcclean2017openfermion, dahlberg2017simulaqron, qengine}.

\section{Exemplifying the Numerical Experiment Design Methodology}
\label{sec:results}

We present the results of two sets of numerical experiments, in accordance with the numerical experiment design methodology introduced in \secref{sec:experiment_method}, utilising discussions, in \secref{sec:preliminaries}, regarding the problem, architecture and simulator to be used. The first considers the 2D-DQS problem, the restricted class of \IQP{} computations presented in \secref{sec:2d-dqs}, and is used to demonstrate the potential of classical simulators as a tool to guide experimental research. In \secref{sec:NQIT noise experiment} and \secref{sec:projections}, where we present results for this problem, we simplify NQIT architectural constraints to focus on the impact of noise. 

We embrace the full complexity of the NQIT architecture in a second numerical experiment presented in \secref{sec:NQIT_experiment}. We restrict a general IQP-MBQC problem seen in \secref{sec:IQP in MBQC} to the NQIT architecture. The hardness of the \IQP{} problem could, in principle, be destroyed by these restrictions and thus further theoretical investigation is required. Here we focus on the effect of architectural constraints on simulations, while the proof of hardness and detailed noise analysis is left for future works. 

While we will reference the simulation details, architectural constraints and figures of merit used in each of the experiments, we note some traits which will be common in all of our experiments.

\paragraph{Simulation} 

To introduce some terminology, each \emph{numerical experiment} consists of several \emph{trials} which are simulations of several different but related circuits. Often a trial will consist of many \emph{runs}, themselves involving several simulations of the same circuit. For example, an experiment might have many trials, each containing a run simulating a probability amplitude for an output of a perfect circuit and several runs each simulating the same output probability amplitude, but with different noisy versions of that circuit.

Indeed, each trial will compare a \emph{perfect run} and possibly several \emph{noisy runs}, which we will identify in each numerical experiment. In particular, in the case of numerical experiments benchmarking the simulator itself the perfect run will be conducted using a brute force simulator while the noisy run will utilise the simulator of our choice. In this case, the noisy simulation is noisy in the sense that the outcomes of the chosen simulator are probabilistic. In the cases where the device is being benchmarked, the perfect run will not consider the architectural noise model, while each noisy runs will.

\paragraph{Constraints} 

Within each numerical experiment we must identify the constraints on the family of circuits we are considering in order to ensure that it is consistent with the philosophy of this paper. In particular, we must ensure that the perfect runs have the necessary theoretical support, for which we will fall back on the \IQP{} hardness results detailed in \secref{sec:IQPintro}. Indeed, in that section we justified considering additive error worst case hardness to be a sufficient support when a demonstration of quantum-advantage is being considered.

A general restriction which is pervasive in our work concerns the degree to which operations can be parallelised in the circuits we consider. While, in theory, \IQP{} circuits are parallel by construction, qubits are physical systems and, in the circuit model, one may be required to apply multiple gates on the same systems. Experimentally it may not be possible to perform these gates simultaneously, even if the gates commute with each other. However, to increase parallelisation of the computation, in our numerical experiments we consider cases of IQP-MBQC where all measurements can be made simultaneously, allowing us to neglect the impact of time based noise during measurement. If we used a less parallel realisation of \IQP{} circuits, it would be prone to the same type and size of noise as a general universal quantum computation and would not be a better candidate for demonstrating quantum-advantage than a universal quantum computation.

Similarly, as discussed in \secref{sec:NQIT_arc}, while the NQIT Q20:20 device is universal, to apply a 2-qubit gate on qubits which belong to ion-traps that are far apart on the 2D lattice, would require many swap gates, each consuming linking qubits. This can result in a large overhead \cite{herbert2018depth, brierley2015efficient, childs2019, cowtan2019qubit} and so a high noise level. Thus, we aim to minimise the number of such gates when deriving our restrictions and we will see that very few swap gates are required for our choices of problems.

\paragraph{Figures of Merit}

To compare perfect runs, which will be justified in their use by the discussion on constraints, with noisy runs, we must consider what figures of merit we will use to judge the quality of those noisy runs. When quantum-advantage is not of concern, for example when benchmarking the classical simulator we use, as is demanded by Part 1 of the numerical experiment design methodology of \secref{sec:experiment_method}, and as we do in \secref{sec:latice_experiment}, the figure of merit will relate to its reliability in producing accurate outcomes. 

\begin{description}
    \item[Statistical test for model closeness] In this case, the output of the simulations are single values of probability amplitudes. A statistical test will be necessary to compare the probability amplitudes from perfect runs to those of the noisy runs. We will use the coefficient of determination to measure the quality of the noisy runs as a model for the perfect runs. This is detailed further in \secref{sec:latice_experiment}.
\end{description}

In the case of simulator benchmarking we compare the probability amplitudes from a brute force simulation to those of the probabilistic simulator, which can be seen as a model of the brute force simulator. We use the same statistical test in \secref{sec:NQIT_experiment} when we simulate restricted instances of the detailed NQIT architecture as we are less concerned by exploring quantum-advantage when the theoretical foundations has been weakened by this restriction. There we will focus on the application of our work to restricted architectures, and study the implications for more general architectures, but find the quantum-advantage motivated figures of merit discussed below to be inappropriate there.

By comparison, when considering the prospect of a device demonstrating a quantum-advantage, the figure of merit will relate to the anticipated usefulness of a larger scale real world implementation of the circuits we are simulating in demonstrating said quantum-advantage. Such a consideration is demanded by Part 2, device benchmarking, and Part 3, guiding future experiments, of the numerical experiment design methodology of \secref{sec:experiment_method}, and is performed in \secref{sec:device benchmarking} and \secref{sec:projections}.

In the case of the simulations of noisy circuits in \secref{sec:device benchmarking} and \secref{sec:projections}, while we do not formally consider their hardness, our measure will be the closeness of the simulated probability amplitudes to the perfect simulations, for which the hardness results of \secref{sec:IQPintro} apply. The theoretical results regarding the hardness of noisy distributions typically concern its \lone{} from the perfect distribution. In our case we do not have access to this information because, as discussed in \secref{sec:class_sim} where the simulator we use is introduced, we access only the amplitudes of a single output, rather than fully characterising the distribution. As such we will often use proxy measures of the \lone{} between perfect and noisy distributions. 

While there are classical simulations which would give us access to a full characterisation of the distribution, here we forgo this option. This is because our aim is to explore the impact of noise at the boundary between what can be simulated classically and what cannot. To do so we have chosen to use a simulator which allows us access to a higher number of qubits than can be implemented experimentally on the NQIT architecture, and than could be implemented using simulators which characterise the full probability distribution. Indeed, the challenge of fully characterising all $2^n$ probabilities of and $n$ qubit probability distribution quickly becomes insurmountable as the number of qubits grows, and certainly becomes increasingly hard as the circuits considered approach the regime of quantum-advantage.

With this in mind, we note the following figure of merit which will be used in \secref{sec:device benchmarking} and \secref{sec:projections}.
\begin{description}

    \item[Accuracy and far from uniformity of noisy runs:] We will consider a numerical experiment to have demonstrated that the current noise values are likely to bring implementations within the reach of classical simulation if trials show either; the noisy probability amplitudes to be within a standard deviation of a uniform distribution amplitude, or greater than one standard deviation from the perfect amplitude. 
    
\end{description}

This condition is reasonable as it asserts that for it to be considered possible for a distribution to demonstrate quantum-advantage it must both have outcomes with probabilities which are far from the uniform distribution value, and that the values of these probabilities are close to the ideal ones. 

These far from uniform points are of great importance for several theoretical reasons. Their existence is shown to be indicative of quantum-advantage \cite{boixo2016characterizing, aaronson2013bosonsampling} while their accuracy is also shown to be vital. For example, studies of the heavy outputs of random circuits \cite{aaronson2016complexity} show that a device could demonstrate quantum-advantage by preserving those probabilities with higher than median value. Indeed other benchmarks of a quantum device's ability to demonstrate quantum-advantage have been build around this realisation \citep{Bishop2017QuantumV}. In addition, measures such as multiplicative error, on which many quantum-advantage statements are based, as discussed in \secref{sec:iqp_thms}, and cross entropy difference, which has also been used to benchmark quantum devices and their demonstration of quantum-advantage \cite{boixo2016characterizing}, are particularly sensitive to the effect noise has on outcomes with small probabilities. In the extreme case, they consider a noisy distribution which does not preserve the probability of outcomes which are not in the support of the ideal distribution to be infinitely far from the ideal distribution.

% This condition would be reasonable The unlikelihood of classical simulation of distributions passing this conditions follows because it ensures that there exists outcome probability values in the noisy distribution which are far from uniform. 

Contradicting this accuracy and far from uniformity statement can therefore be seen as an indication, but not proof, of the ability to demonstrate quantum-advantage in the setting being considered. Indeed we will consider a demonstration of quantum-advantage to be more likely if the noisy distribution more often contradict the statement, which is to say that there are more outcomes with probability both far from the uniform value, and close to the ideal value. While there is an upper bound to the number of such outputs, namely the number of outputs with probability far from the uniform value in the ideal distribution, this figure of merit provides a lower bound on what must be achieved for a demonstration of quantum-advantage to be deemed likely.

% We encourage this form of the distribution as this separation from the uniform distribution is thought to be an indication of quantum-advantage , and that those values are close to their ideal value. In other words this condition ensures that those probabilities which are far from the uniform probability value in the ideal distribution maintain this property in the noisy distribution.

This accuracy and far from uniformity measure also implies that values close to uniform ones in the ideal distribution remain so in the noisy distribution. This follows as such values would otherwise be `far from the perfect amplitude'. However, in many cases noise has the effect of bringing probability values close to the uniform distribution and so little information about the effect of noise can be obtained from these outputs as they will be little changed. For the noise types listed for the NQIT device in \secref{sec:noise} it is the case that they result in a convergence of the output distribution to the uniform one. While this is not the case for noise channels such as amplitude damping, these errors would also be captured by this figure of merit as it would have the effect of decreasing the probability of likely outputs towards the uniform distribution value. Further, it would be impossible to distinguish close to uniform values which have been achieved through accurate reproduction of the ideal distribution and those which have been achieved through a naive approximation by a uniform distribution. While it is true that these points are of value to the form of the distribution as discussed in \secref{sec:iqp_thms}, as we cannot make this distinction we do not include them in our analysis. By isolating outcomes which have far from the uniform probability in the ideal distribution we obtain the additional advantage of being able to limit the outputs which we must study in our experiments, allowing us to run larger circuits as a trade off.

Due to the anticoncentration property of \IQP{} distributions \cite{Bremner2015}, this might result in considerable filtering of our simulations. However, while results about the quantum-advantage of \IQP{} circuits, seen in \secref{sec:IQPintro} and utilised in the constraints section of each numerical experiment, assure us that the circuits we consider cannot, in general, be simulated classically, this filtering removes output amplitudes which could trivially be simulated using a uniform distribution.

That contradicting this condition implies there are points which are further than one standard deviation from the uniform distribution is valuable in that it ensures points which are far from uniform in the ideal distribution reliably remain this way in the noisy one. This is vital for the reasons stated above. That they be within one standard deviation of the ideal is not strictly necessary as quantum-advantage statements typically allow for a constant deviation of the whole distribution. However, as we are unable calculate the deviation of the whole distribution, this again seems like a reasonable proxy. In fact, as we will see, in \secref{sec:TV proxy} there is a relationship between this measure and more direct proxies for the \lone{}.

% In the case of those outputs the measure corresponds to the statement that the probabilities must be close to the ideal with standard deviation not encompassing the uniform distribution. 

% This restriction to the consideration of only far from uniform amplitude outputs corresponds to theoretical results in several different ways.  As a result we will filter our results here to account for this priority.

% Because of the lack of directly applicable theoretical results for this measure, it can only reasonably be used to discount the possibility of a demonstration of quantum-advantage, rather than imply the existence of one. 

However, as we focus on single amplitudes, it may be that this is a strong metric. While it is shown to be hard for a classical device to sample from the output distribution of arbitrary IQP circuits, which are subject to noise, up to a small relative error in each probability \cite{fujii2014computational}, this is possible up to \lone{} \cite{Bremner2017achievingsupremacy}.

% This condition implies that the corresponding perfect runs must themselves be far from the uniform distribution probability. As such, in the numerical experiments that follow, we will often ignore those trials with a perfect run that produces a value which is close the that of the uniform distribution. We will detail in each section what our measure of closeness will be.

The previous two figures of merit have the advantage that they are the best utilisation of the simulator that we have chosen to use. In particular, they extract a significant amount of information from the single probability values which we have access to. That being said, as we mentioned before, theoretical results often refer to global properties of the probability distributions. The following figure of merit addresses this disparity.
    
\begin{description}
    \item[Close in \lone{}:] When the circuit considered in a numerical experiment is considered to not be unlikely to demonstrate quantum-advantage, as defined in the above condition, we will consider the closeness of the noisy and perfect runs using proxies for the \lone{}.
\end{description}

Because of the relationship between this figure of merit and the theoretical results about \IQP{} in \secref{sec:iqp_thms} and, in particular, 2D-DQS in \secref{sec:2d-dqs}, this figure of merit can more reasonably be expected to be a predictor of demonstrations of quantum-advantage than in the previous case. Once again we will often refer to the relative likelihood of a demonstration of quantum advantage between noise settings as measured by the degree of improvement in the \lone{}. In this case we have the additional benefit of being able to study the closeness of the measured value of the \lone{} to the value specified in the relevant theoretical results of \secref{sec:preliminaries}, although this proxy of the \lone{} does not allow us to make formal claims.

By encapsulating results related to those far from uniform outcome probability values and proxies for the \lone{} we cover a diverse set of theoretical results. We believe there is great value in this diversified approach and as such we will combine both the accuracy and far from uniformity condition and the close in \lone{} condition throughout our work.

For each numerical experiment we will use considerations of the simulation method, the constraints of the architecture, and the appropriate figures of merit to draw conclusions pertaining to the goals of this paper.

\subsection{Part 1 : Simulator Benchmarking}
\label{sec:latice_experiment}

As we outlined in \secref{sec:experiment_method}, Part 1 of the numerical experiment builds confidence in our simulator by comparing the outputs to a brute-force simulation. Here we detail the numerical experiment used to do so.

\paragraph{Constraints} 

Here we will not consider the specifics of the architectural noise as we are measuring the impact of using a probabilistic simulator as compared to a brute-force one. Moreover, it is sufficient to benchmark the probabilistic simulator by comparing the outputs to those of a brute-force simulation for a \emph{general} IQP-MBQC problem of \secref{sec:IQP in MBQC}. We do not restrict to a particular architecture here but the generality we utilise ensures the functioning of the simulator for restricted instances which we explore later.

\paragraph{Simulation}

As described in \apref{app:IQPMBQC brute}, during each trial we will generate a random instance of the general IQP-MBQC problem of \secref{sec:IQP in MBQC}, and simulate the circuit to obtain the probability of measuring the $\ket{0^n}$ state. The randomly generated circuits will have between $5$ and $12$ qubits, and between $5$ and $15$ $T$ gates. In the case of the perfect run, the solution will be obtained by using the brute force simulator, while in the case of a noisy run it will be solved by taking the mean of several simulations using the probabilistic simulator of \secref{sec:class_sim}. Together these two runs constitute a trial. The resulting values for the runs in each trial are then compared to calculate the coefficient of determination as described in the figures of merit section.

\begin{figure}
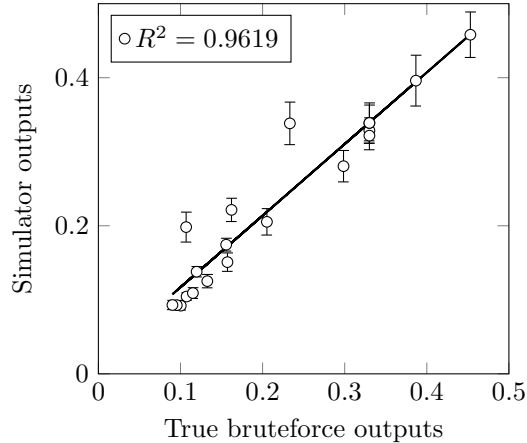

    \centering
    \myinclude{0.5}{0.425}{XprogramBruteforceVsFast}
    \caption{Comparison between brute-force outputs and probabilistic simulator outputs when calculating the probability of measuring the $\ket{0}^{n}$ state for 20 random $X$-programs. Each point indicates the mean probability of measuring the $\ket{0}^{n}$ state for one fixed $X$-program according to the simulator, with the error bars indicating one standard deviation in the probabilistic simulator's output. The number of qubits is in the range [5,12] and the $T$-gate count is in the range [5,15]. Details of this simulation can be found in \apref{app:IQPMBQC}. Strong correlation is observed with $R^2 = 0.9619$. Here, unlike in later plots, the axis are not scaled as the probabilities are of a reasonable magnitude due to the smaller circuit sizes.}
    \label{fig:XprogramBruteforceVsFast}
\end{figure}

As discussed, while the brute-force simulation is deterministic, the simulator of \secref{sec:class_sim} which we are testing against it is probabilistic. As such, each noisy run will consist of calculating the given probability distribution many times, and averaging. The mean and standard deviation are plotted in \figref{fig:XprogramBruteforceVsFast}.

Here it is sufficient to consider only the probability of measuring the state $\ket{0^n}$ as no additional error is added by measuring other states. As measuring other basis states requires only the appropriate $X$ gates, which can be applied deterministically by the simulator of \secref{sec:class_sim}, unlike $T$ gates which are applied probabilistically, no additional error will result from considering only the $\ket{0^n}$ state.

\paragraph{Figures of Merit}

The measure we will use to compare the perfect and noisy runs is the coefficient of determination, which can be said to measure the correlation between the outputs of a model and those from its target. Given outputs $m_i$ from a model, and the corresponding target outputs $d_i$, with mean $\bar{d}$, the coefficient of determination is calculated using \eqref{equ:coef of det}. In \eqref{equ:coef of det}, $r=\sum_i \left( d_i - m_i \right)^2$ is the residual sum of squares and $v = \sum_i \left( d_i - \bar{d} \right)^2$ is the total sum of squares.
\begin{equation}
    \label{equ:coef of det}
    R^2 = 1 - \frac{r}{v}
\end{equation}
In our case the model is the simulator of \secref{sec:class_sim} and the target is the brute force simulation. The data, $d_i$ and $m_i$, are the values for the amplitudes of the $\ket{0^n}$ state produced by the brute-force and probabilistic simulator, respectively, during the $i$\textsuperscript{th} trial.

\paragraph{Conclusion}

Results in \figref{fig:XprogramBruteforceVsFast} show that the average of the simulator outputs exhibit strong correlation with the true values from a brute force simulation, giving a coefficient of determination $R^2 = 0.9619$. As such we can have confidence in our choice of simulator for the problems we will tackle in the following sections.

\subsection{Part 2 : Device Benchmarking}
\label{sec:device benchmarking}

Continuing to follow the method of \secref{sec:experiment_method}, Part 2 of each numerical experiment is to impose the constraints that come from the experimental system used. In the following we restrict, with differing degrees of strictness, problems previously mentioned, to the NQIT architecture.

\subsubsection{NQIT Noise Restricted 2D-DQS}
\label{sec:NQIT noise experiment}

We consider Architecture I from \cite{Bermejo-Vega2017} as discussed in \secref{sec:2d-dqs} and constrain it according to the noise of the NQIT machine as listed in \secref{sec:noise}. For simplicity, we will use a modified version of the NQIT architectural restraints of \secref{sec:NQIT_arc}.

\paragraph{Constraints} 

The 2D-DQS problem has been designed for networked architectures and, with some simple adaptations, it can coincide with NQIT's device. In particular, by making the simplifying assumption that we use a single logical qubit per ion-trap\footnote{Using more qubits per ion-trap could be possible, but then the connectivity of qubits would not be identical to that of the problem considered. Since in this example we focus on the issue of noise, we make this assumption and let non-trivial architectural constraints be considered in the next numerical experiment.} we can map every grid vertex onto a single ion trap. One may then look to \figref{fig:archetecture I} and \figref{fig:NQIT arc} to understand that the 2D-DQS problem can be easily overlaid onto the NQIT architecture, which also permits the necessary measurements, state preparations, and single and 2-qubit gates.

As the adapted NQIT architectural restrains, detailed above, adhere to those required for the 2D-DQS problem seen in \algref{jens}, the worst case additive error hardness result of the 2D-DQS problem, as seen in \theref{2D-DQS hardness}, applies. While we have agreed that this setting constitutes one that is worthy of investigation, as the noise levels are independent for each qubit and not dependent on the problem size, the additive error permitted by \theref{2D-DQS hardness} is likely exceeded. Hence, we would expect that in the noisy case the distribution becomes far from the perfect one and for the advantage to diminish.

\paragraph{Simulation} 

We consider $4\times 5$ grids, modelling $20$ ion traps in total, and use them to perform \algref{jens}. \algref{jens} requires, on average, half as many $T$ gates as qubits; in this case $10$ and $20$ respectively. Details of the numerical specifics of the experiments can be found in \ref{sec:appendixJens}. Here it suffices to say that we use four steps to generate the entangled 2D cluster. The number of steps plays a role in the amount of noise as it determines the duration of the computation and thus the decoherence time we consider.

We perform $20$ trials, each concerning one perfect circuit and a random output string. For each trial there are $20$ noisy runs, each with their own random noisy version of the trial's circuit. This random noisy version of the perfect circuit is generated by considering the noise type and strength of the experiment as described in \ref{sec:appendixJens}. We simulate all $21$ circuits $20$ times, calculating the mean probability of measuring the corresponding bit string in each case. We will then take the mean and standard deviation of the noisy runs.

While, as noted in \cite{Bravyi2016}, simulations of up to $40$ qubits and $50$ $T$ gates is possible using this simulator, as is also noted in that work, doing so takes several hours. In our case we simulate $20$ trials, each with $21$ runs and $20$ simulations per run and so we restrict the number of qubits and $T$ gates to a more manageable amount. Later in this work we go further and perform many thousands of simulations in each numerical experiment, justifying our restriction.

\paragraph{Figure of Merit} 

For this numerical experiment we will utilise the `accuracy and far from uniformity of noisy runs' condition from the introduction to \secref{sec:results}. In particular, we will consider a perfect run to be far from uniform when it is either greater than twice the uniform value, or less than half. In this way we will identify if the noise level reveals that, as we expect, the potential for a demonstration of quantum-advantage should be dismissed, rather than if one could be achieved.

\paragraph{Conclusion} 

\begin{figure}[ht]
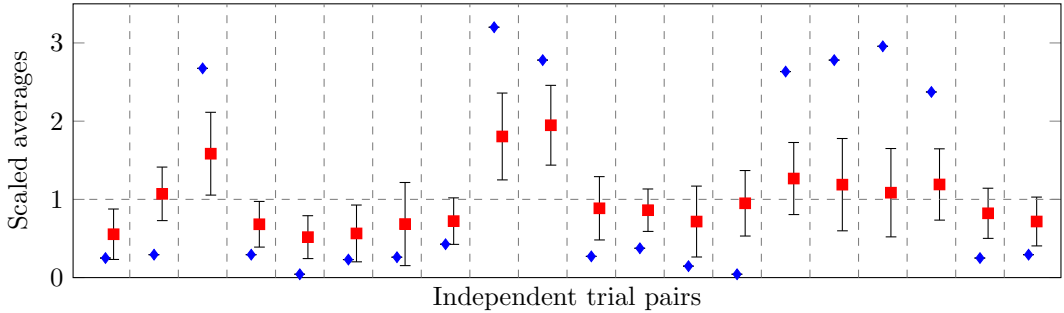

	\centering
	\myinclude{1}{0.3}{jensFastVsFastNoisy}
	\caption{Comparison between ideal and noisy circuit results for a $4\times 5$ ion trap grid. The results referenced by this plot are the probability of measuring a randomly chosen output string, where each trial has a different initial 2D-DQS circuit, and different output string. Every consecutive pair is one trial and contains the perfect run (blue diamond), and the mean of the noisy runs (red square). The error bars indicate one standard deviation of the noisy runs. The means and standard deviations for each trial have been normalised by the uniform distribution (dotted horizontal line).} 
	\label{fig:jensFastVsFastNoisy}
\end{figure}

The results are shown in \figref{fig:jensFastVsFastNoisy} where we have plotted the value for the perfect run, and the mean value for the noisy runs. As expected, including noise at the levels of the NQIT device leads to an outcome probability that is between the ideal and the totally random output. However in most cases the noise that we include leads to a result within one standard deviation of the uniform distribution, or greater than one standard deviation from the perfect run. Referring to our figures of merit, we regard this to be a sign that the scheme is unsatisfactory for demonstrating quantum-advantage with NQIT noise at its current levels.

In \secref{sec:projections} we use the simulator as a tool to investigate which of the aspects of our noise model are the main sources of this failure. Our intention is to direct subsequent experimental  and theoretical research towards diminishing this source, potentially leading to a quicker implementation of quantum-advantage experiments.

To form a complete picture, and to benchmark the device's performance when implementing these problems, we must compare our numerical experiments with actual experiments. This work concerns only numerical experiments, while in the future we plan to collaborate with experimental groups to provide these benchmarks.

\subsubsection{NQIT Noise and Architecture Restricted IQP-MBQC}
\label{sec:NQIT_experiment}

The second numerical experiment we perform takes the general IQP-MBQC of \secref{sec:IQP in MBQC} and imposes constraints equivalent to the architecture of NQIT. We consider the case where each ion-trap has multiple logical qubits, as discussed in \secref{sec:NQIT_arc}. Moreover, we restrict to \IQP{} instances involving gates acting on qubits belonging to neighbouring ion-traps so as to lower the circuit depth as much as possible. 

\paragraph{Constraints}

In principle different gates of an $X$-program may act on any subset of qubits, or in the MBQC model, the gate qubits may be entangled with any subset of the application qubits. This is not realistically achieved in the NQIT setting, where qubits belonging in different ion-traps cannot be connected arbitrarily with qubits of other ion-traps. Since NQIT admits universal quantum computation, one could achieve arbitrary connectivity by using swaps between the qubits. However, by doing these swaps the advantage of smaller waiting times offered by \IQP{} is destroyed. We will thus impose conditions on the connectivity, limiting the class of problems we use.

We have assumed that each ion-trap has $K=20$ physical qubits, of which $10$ are dedicated to entanglement distillation, leaving $K'=10$ for use in computation. As discussed in \secref{sec:noise}, this allows us to fix the noise of two-qubits gates to be constant, whether it involves qubits in the same or neighbouring ion-traps. This does not apply to the waiting time, and thus decoherence, which is greater in the case of gates involving qubits in different ion-traps.  

We will choose the minimum links between different ion-traps (while maintaining full connectivity within each trap). This means a 1 dimensional configuration of ion-traps\footnote{We could consider the 2 dimensional case too, as in the first numerical experiment, but our choice is the simplest and within reach of our classical simulator. A 2 dimensional configuration would require a larger number of traps, which is outside of our simulation capabilities.}. This, in itself, might not be a big restriction, since even considering two-qubit gates that act on nearest neighbour qubits only, as shown by \theref{2d iqp hard}, is still believed to be a hard problem. However, this configuration, while it is not 1 dimensional as far as the qubits are concerned, is still likely to admit a classical efficient simulation based on tensor networks and matrix product states \cite{MPS}. Since our purpose in this section is to illustrate how to implement architecture constraints, the issue of classical hardness in comparison to the best classical methods, is not crucial. Indeed it is likely that reasonable predictions can be made about the impact of noise on the 2 dimensional architecture designed by NQIT, outputs from which are less likely to be reproducible on a classical computer, using results from these 1 dimensional simulations. In contrast, in the first numerical experiment, there \emph{is} a complexity-theoretic proof of hardness. 

In IQP-MBQC, applying gates between application qubits corresponds to entangling them with the same gate qubit. In the case that the application qubits belong to different ion-traps, the gate is applied using teleportation, with the help of entanglement links distilled between neighbouring ion-traps. Protocol \ref{alg:entangleTwoTraps} shows how to achieve this using only one entanglement link between the two ion-traps. Distilling entanglement between multiple traps takes a longer time, which is why we restricted our attention to $X$-programs that involve gates with qubits in at most two ion-traps. 

\begin{algorithm}[ht]

\begin{algorithmic}[1]
	\Function{EntangleTwoTraps}{$p$, $g$, $c$, $l_1$, $l_2$, $Q_1$, $Q_2$}
	\ForAll{$q \in Q_1 : p(q) = 1$}
	\State $C Z$ ($g$, $q$)
	\EndFor
	\State $C Z$ ($g$, $l_1$)
	\State Distil a Bell pair between $l_1$ and $l_2$
	\State Bell measurement on ($g$, $l_1$) which teleports g to $l_2$
	\State SWAP ($c$, $l_2$)
	\ForAll{$q \in Q_2 : p(q) = 1$}
	\State $C Z$ ($c$, $q$)
	\EndFor
	\EndFunction
\end{algorithmic}

\caption{This algorithm constructs part of the resource state for a given gate qubit $g$ in trap 1 according to its corresponding row $p$ of the $X$-program $\mathbf{Q}$. $Q_1$ is the set of all qubits in cell 1 with $g, l_1 \in Q_1$. Analogously, $c, l_2 \in Q_2$. $c$ is the qubit that will eventually be used for measurement after $g$'s value is teleported there.}

\label{alg:entangleTwoTraps}

\end{algorithm}

In this setting, we have each ion-trap being connected by entanglement links to two neighbouring ion-traps. Each ion-trap has one gate qubit ($g$ in \algref{entangleTwoTraps}) and one qubit reserved to receive the gate qubit coming from it's neighbour ($c$ in \algref{entangleTwoTraps}). This leaves 8 application qubits. This entanglement structure can be achieved in two time-steps. First, all ion-traps at odd positions use their entanglement links to teleport the qubit required using Protocol \ref{alg:entangleTwoTraps}. This is repeated for all even positions. This two-step process is shown schematically in \eqref{eq:trapsTwoSteps}.

\begin{equation} 
\label{eq:trapsTwoSteps}
\newcommand{\tikzmark}[1]{\tikz[overlay,remember picture] \node (#1) {};}
\tikzset{square arrow/.style={to path={-- ++(0,-.25) -| (\tikztotarget)}}}
1\tikzmark{a} \quad 2\tikzmark{b} \quad 3\tikzmark{c} \quad 4\tikzmark{d} \: \quad \rightarrow \quad 1\tikzmark{a1} - 2\tikzmark{b1} \quad 3\tikzmark{c1} - 4\tikzmark{d1} \: \quad \rightarrow \quad 1\tikzmark{a2} - 2\tikzmark{b2} - 3\tikzmark{c2} - 4\tikzmark{d2} \:
\tikz[overlay,remember picture]
{\draw[->,square arrow] (a.south) to (b.south);\draw[->,square arrow] (c.south) to (d.south); \draw[->,square arrow] (b1.south) to (c1.south);}
\end{equation}

\vspace{1em}

\begin{figure}[ht]
	\centering
	\begin{tikzpicture}[-latex ,auto ,node distance =5 mm and 5mm, on grid, semithick ,state/.style ={ very thick, circle, draw, minimum width = 0.3mm}]
    	\node[state] (a11) {}; 
	    \node[state] (a12) [below =of a11] {};
	    \node[state] (a13) [below =of a12] {};
	    \node[state] (a21) [right =of a11] {}; 
    	\node[state] (a22) [below =of a21] {};
    	\node[state] (a23) [below =of a22] {};
    	\node[state] (a31) [right =of a21] {}; 
    	\node[state] (a32) [below =of a31] {};
    	\node[state] (a33) [below =of a32] {};
		
    	\node[state] (b11) [right =of a11, xshift=3cm]{}; 
    	\node[state] (b12) [below =of b11] {};
    	\node[state] (b13) [below =of b12] {};
    	\node[state] (b21) [right =of b11] {}; 
    	\node[state] (b22) [below =of b21] {};
    	\node[state] (b23) [below =of b22] {};
    	\node[state] (b31) [right =of b21] {}; 
    	\node[state] (b32) [below =of b31] {};
    	\node[state] (b33) [below =of b32] {};
		
    	\node[state] (c11) [right =of b11, xshift=3cm]{}; 
    	\node[state] (c12) [below =of c11] {};
    	\node[state] (c13) [below =of c12] {};
    	\node[state] (c21) [right =of c11] {}; 
    	\node[state] (c22) [below =of c21] {};
    	\node[state] (c23) [below =of c22] {};
    	\node[state] (c31) [right =of c21] {}; 
    	\node[state] (c32) [below =of c31] {};
    	\node[state] (c33) [below =of c32] {};
		
    	\node (ga) [below =of a22,yshift=3cm] {};
    	\node[state, dotted] (gb) [below =of a22,yshift=3cm] {};
    	\node[state] (gb) [below =of b22,yshift=3cm] {};
    	\node[state] (gc) [below =of c22,yshift=3cm] {};
    	
    	\path[-,very thick] (gb) edge (a11);
    	\path[-,very thick] (gb) edge (a13);
    	\path[-,very thick] (gb) edge (a22);
    	\path[-,very thick] (gb) edge (a31);
    	\path[-,very thick] (gb) edge (b21);
    	\path[-,very thick] (gb) edge (b13);
    	\path[-,very thick] (gb) edge (b33);
    	
	    \path[-,very thick] (gc) edge (b21);
    	\path[-,very thick] (gc) edge (b23);
    	\path[-,very thick] (gc) edge (b31);
    	\path[-,very thick] (gc) edge (c11);
    	\path[-,very thick] (gc) edge (c22);
    	\path[-,very thick] (gc) edge (c33);
    	\path[-,very thick] (gc) edge (c12);
    	\path[-,very thick] (gc) edge (c31);
    		
    	\draw[very thick, dashed,-] (-1,1) -- (9,1);

	\draw[very thick, dashed, rounded corners = 20pt] (-0.5,-1.5) rectangle (1.5,2.5);
	\draw[very thick, dashed, rounded corners = 20pt] (3,-1.5) rectangle (5,2.5);
	\draw[very thick, dashed, rounded corners = 20pt] (6.5,-1.5) rectangle (8.5,2.5);
	
	\end{tikzpicture}
	\caption{An example of a restricted MBQC pattern for 3 traps, where application qubits are on the bottom and gate qubits are on the top. Gate qubits are still physically in the cells with the application ones, although they are separated by a dotted line here for clarity. We have one gate qubit for every two neighbouring cells, with considerations made for boundary cases. Once a gate qubit is entangled in its native trap it is moved. There is one less gate qubit than the number of traps so that each is entangled to two traps. The dotted gate qubit indicates a location which has been vacated when the gate qubits move between traps. The reader may wish to return to \figref{fig:NQIT arc} where, like here, the dashed bubbles indicate individual ion traps with a single qubit in each acting between them.}
	\label{fig:XprogramExampleRestrictedPattern}
\end{figure}
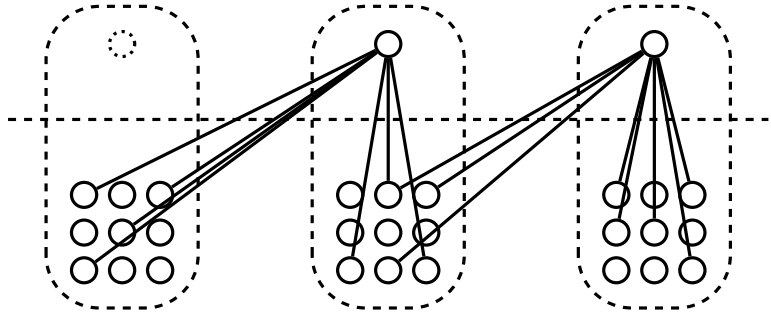

With these restrictions $X$-programs can be mapped to NQIT's architecture. An example of an MBQC graph for such restricted instances is given in \figref{fig:XprogramExampleRestrictedPattern}.

\paragraph{Simulations} 

A full description of the simulation procedure can be seen in \apref{app:IQPMBQC NQIT}. In summary, we let each gate qubit act on a random subset of the application qubits in its own ion-trap before, after being teleported, acting on a random subset of the qubits in the next ion-trap. We performed $20$ trial, each involving a randomly generated circuit of the form described above, along with a random output string. Each trial has one noisy and one perfect run. A perfect run involves simulating the perfect circuit several times and calculating the mean probability of measuring the selected output string. A noisy run is equivalent but with a random noisy instance of the circuit. 

In this case, at their largest, we simulate significantly more qubits than in the previous and following sections. The largest circuit we simulate has $12 \times 8$ qubits but still only $10$ $T$ gates on average. This is because we have limited the probability that a $T$ gate will be required, which corresponds, as discussed in \apref{app:IQPMBQC NQIT}, to limiting the probability of creating connections between the gate and application qubits. As the computation time grows exponentially with the number of $T$ gates, and polynomially in the number of qubits, we can afford this increase in the qubit count.

\paragraph{Figure of Merit} 

In this case, as we expect that the architectural restrictions used will make a demonstration of quantum-advantage using this scheme unlikely, we will not consider the figures of merit as in \secref{sec:NQIT noise experiment}. Instead we again consider the coefficient of determination as in \secref{sec:latice_experiment} to establish the impact of noise models more broadly. Here the model outputs $m_i$ are the probability amplitudes from the noisy run, while the target outputs $d_i$ are those from the perfect run.

\paragraph{Conclusion} 

We compared the two means of each run to calculate the coefficient of determination. In the case of the maximum system ($12$ ion-traps, with $8$ application qubits each) we noticed that, with the existing level of noise, the results corrupt fully the output leading to $R^2\approx 0$. We then ran similar experiments for smaller instances. Lowering the number of qubits, we observed that the $R^2$ value was increasing but still remained extremely low with NQIT noise level. Decreasing the size yielded the following results ($a \times b$ means $a$ ion-traps with $b$ application qubits per trap):

\begin{center}
	\begin{tabular}{ c | c c c c }
 		$a \times b$	& $12 \times 8$	& $9 \times 8$	& $4 \times 8$	& $4 \times 2$ \\ 
 		$R^{2}$		& $0.0086$ 	& $0.0237$ 	& $0.0333$	& $0.5561$
	\end{tabular}
\end{center}

These $R^2$ values, far below one, indicate that even for small system sizes, the noise is too high and there is little correlation between the perfect and noisy runs. For this reason, and because theoretical results about quantum-advantage in this case are not as strong, in the subsequent section where we examine the effects of varying noise, we restricted attention to the numerical experiment of \secref{sec:latice_experiment} only and do not proceed to Part 3 of the numerical experiments in this case.

\subsection{Part 3 : Guiding Future Experiments Using NQIT Noise Restricted 2D-DQS}
\label{sec:projections}

To identify the main sources of error in the numerical experiment of \secref{sec:NQIT noise experiment} we run experiments with varying noise levels. In this section, the protocol we implement will be the 2D-DQS of \secref{sec:2d-dqs} as detailed in \ref{sec:appendixJens}. We group the different noise types of \secref{sec:noise} together and identify which contributes most to the corruption of the perfect output. We then ``fine-grain'' further by considering the different types of noise within that group. Once we have identified the main source of error, we will explore how the potential for a demonstration of quantum-advantage is affected by reducing this noise, both by known error correction techniques, and hypothetical proposals. 

In these numerical experiments will use the same constraints and simulation design as in the first 2D-DQS simulations of \secref{sec:NQIT noise experiment}. The difference here is the noise model used. In particular, we will be comparing random single output probabilities. We will also use the same `accuracy and far from uniformity of noisy runs' figure of merit as in \secref{sec:NQIT noise experiment} in order to identify when a demonstration of quantum-advantage is unlikely. As we identify cases where such a demonstration is not unlikely, we will explore proxy measures for the \lone{} and relate these measures back to the theoretical results, in \secref{sec:2d-dqs}, on the conditions for a demonstration of quantum-advantage using the 2D-DQS protocol.

\subsubsection{Operation-Based Verses Time-Based Noise}
\label{sec:op vs time}

\begin{figure}
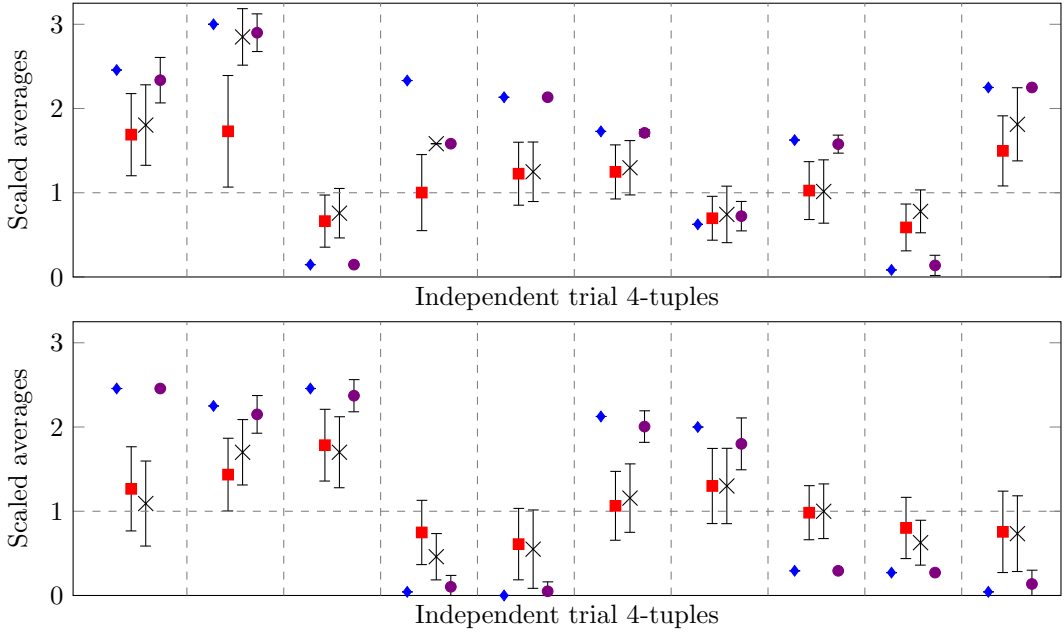

	\centering
	\myinclude{1}{0.3}{jensFastVsFastNoisyVarious_1}
	\myinclude{1}{0.3}{jensFastVsFastNoisyVarious_2}
	\caption{Results including either only gate based noise or only time based noise rates for a $4 \times 5$ ion trap grid. The results referenced by this plot are the probability of measuring a randomly chosen output string, where each trial has a different initial 2D-DQS circuit, and different output string. Every independent trial is described by a 4-tuple of a perfect run (no noise) (blue diamond), the mean of 20 noisy runs (red square), the mean of 20 only time-based rates noisy runs (grey cross) and the mean of 20 only gate rates noisy runs (violet circle). The error bars show one standard deviation. The means and standard deviations have been normalised by the respective uniform distribution (dotted horizontal line).}
	\label{fig:jensFastVsFastNoisyVarious2}
\end{figure}

At the coarsest level of detail, we group time-based noise (depolarising and dephasing) together, and operation-based noise (preparation, measurement, single and two qubit gates, including the noise during distillation) together. In each run we eliminate either the time-based noise or operation-based noise, while keeping the other at the same level as in NQIT's device. Result for the behaviour of outputs with far from uniform probability in the ideal output distribution can be seen in \figref{fig:jensFastVsFastNoisyVarious2}. 

We can see that the largest contribution to the corruption of the output appears to be from the time-based noise. When we were exploring candidates for demonstrating quantum-advantage, we mentioned that time based noise is frequently a major issue. This motivated us to consider \IQP{} and here our results justify this choice.

With reference to our figures of merit, including only time-based noise almost always brings the output probability of the bit string in noisy runs to within one standard deviation of the uniform value, or greater than one standard deviation away from the perfect run amplitude value. As such we conclude that it is a significant obstacle to demonstrating quantum-advantage. On the other hand, as the randomly selected bit string amplitude, when only gate based noise is considered, is in all but one case within one standard deviation of the perfect run, and further than one standard deviation from the uniform distribution value, we do not immediately conclude that it is a significant obstacle. 

Below the reader can see values for a proxy for the \lone{} between the ideal and noisy distributions for the noise levels discussed above, calculated as follows. Here a trial consists of an ideal run, measuring the probability of a random output of a random 2D-DQS circuit of the form discussed in \secref{sec:2d-dqs}, and $20$ noisy runs for each noise type, considering noisy versions of the ideal circuit. The average difference between the noisy and ideal runs within each trial are themselves averaged to give a proxy for the \lone{}, once scaled by the uniform distribution. Each run is itself the average of $20$ simulations of the same circuit.

A similar pattern is seen in this data as was identified in the study of single outputs; namely that the largest contribution to the deviation of the noisy distribution from the ideal is a result of the time based noise.
\begin{center}
    \begin{tabular}{ c|c|c }
        full noise levels & only time base noise & only gate based noise \\ 
        \hline
        0.286316488 & 0.276119941 & 0.033008605 \\ 
\end{tabular}
\end{center}

As discussed in \secref{sec:results} our analysis of both far from uniform outputs and the \lone{} lead us to regard a system with reduced time-based noise as relatively more likely to demonstrate quantum advantage than a system with reduced gate-based noise. In this case, removing the time based noise results in a value below the $\frac{1}{22}$ specified in \theref{2D-DQS hardness} suggesting that a demonstration of quantum-advantage may be possible here. However, we hope to identify the main source of error more precisely, and as such we continue to explore the reduction of time-based noise.

\subsubsection{Depolarising Verses Dephasing Noise}

\begin{figure}
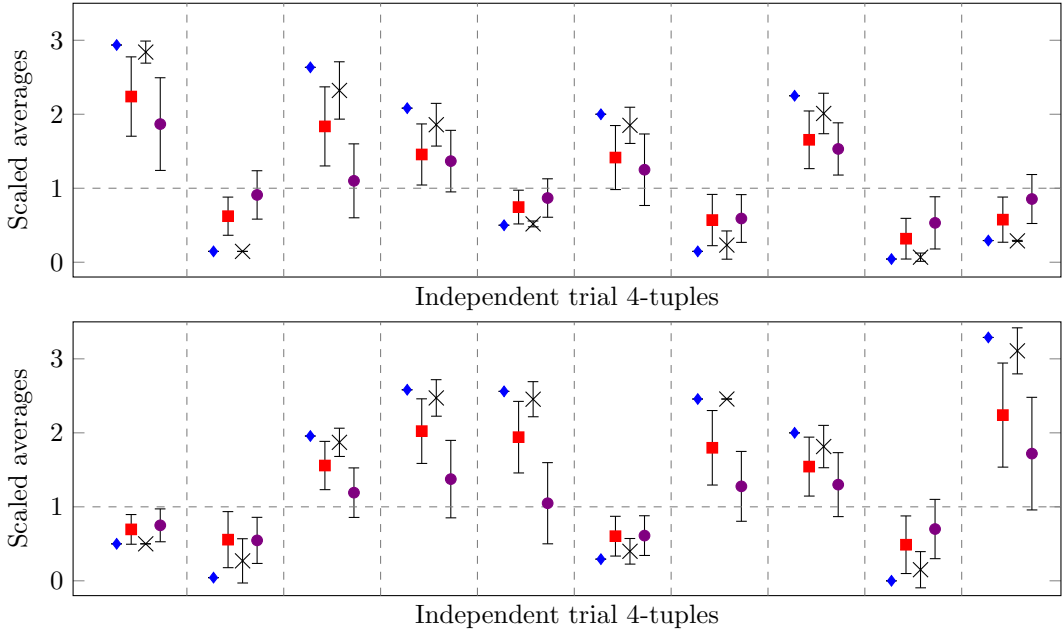

	\centering
	\myinclude{1}{0.3}{jensFastVsFastNoisyVariousTimeBased_1}
	\myinclude{1}{0.3}{jensFastVsFastNoisyVariousTimeBased_2}
	\caption{Results including either only dephasing or only depolarising noise rates for a $4\times 5$ ion trap grid. The results referenced by this plot are the probability of measuring a randomly chosen output string, where each trial has a different initial 2D-DQS circuit, and different output string. Every independent trial is described by a 4-tuple of a perfect run (no noise) (blue diamond), the mean of 20 noisy runs (red square), the mean of 20 only depolarising rates noisy runs (grey cross) and the mean of 20 only dephasing rates noisy runs (violet circle). The error bars show one standard deviation. The means and standard deviations have been normalised by the respective uniform distribution (dotted horizontal line).}
	\label{fig:jensFastVsFastNoisyVarious2timebased}
\end{figure}

We now look more closely at the time-based noise and consider separately the contribution from dephasing noise and from depolarising noise. The results for outputs with far from uniform probability in the ideal output distribution are seen in \figref{fig:jensFastVsFastNoisyVarious2timebased}.

In this case, the amplitudes produced by runs considering only dephasing noise are always either within one standard deviation of the uniform distribution, or greater than one standard deviation from the perfect run. By comparison the runs considering only depolarising errors are always within one standard deviation of the the perfect run, and greater than one standard deviation from the uniform distribution output. 

Below the reader will again find the same proxy for the \lone{} between the ideal and noisy distributions as discussed in \secref{sec:op vs time}, but for the noise levels considered in this section. A similar pattern is seen in this data as was identified when considering the accuracy and far from uniformity figure of merit above; namely that the largest contribution to the deviation of the noisy distribution from the ideal is a result of the dephasing noise.
\begin{center}
    \begin{tabular}{ c|c|c } 
        full noise levels & only depolarising noise & only dephasing noise \\ 
        \hline
        0.433746955 & 0.111777366 & 0.4555678 \\ 
    \end{tabular}
\end{center}

As as a result of the analysis of these two figures of merit, we identify dephasing error as a relatively larger obstacle to a demonstration of quantum-advantage than depolarising noise.

\subsubsection{The Impact of Noise Reduction by Error Correction}

\begin{figure}
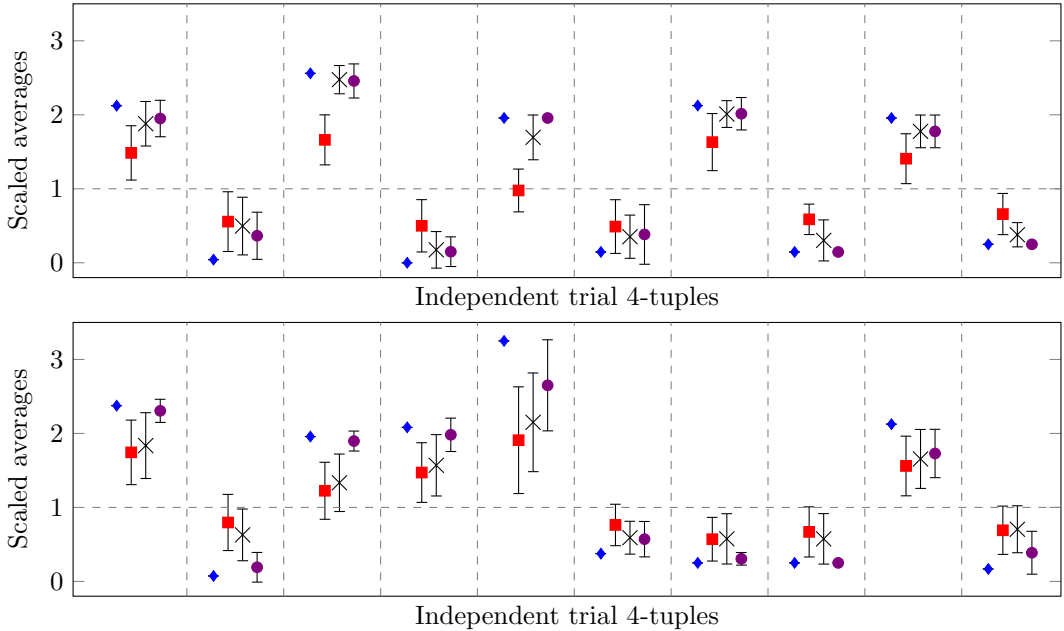

	\centering
	\myinclude{1}{0.3}{jensFastVsFastNoisyVariousDeph_1}
	\myinclude{1}{0.3}{jensFastVsFastNoisyVariousDeph_2}
	\caption{Results including reduced dephasing noise rates for a $4\times 5$ ion trap grid. The results referenced by this plot are the probability of measuring a randomly chosen output string, where each trial has a different initial 2D-DQS circuit, and different output string. Every independent trial is described by a 4-tuple of a perfect run (no noise) (blue diamond), the mean of 20 noisy runs (red square), the mean of 20 dephasing rates reduced by repetition code noisy runs (grey cross) and the mean of 20 no dephasing rates noisy runs (violet circle). The error bars show one standard deviation while. The means and standard deviations have been normalised by the respective uniform distribution (dotted horizontal line).}
	\label{fig:jensFastVsFastNoisyVarious2Deph}
\end{figure}

Having identified the main obstacle to a demonstration of quantum-advantage to be dephasing errors, we examine the effect that reducing this type of noise would have. Concretely, one could introduce a phase-flip code\footnote{This idea was suggested earlier by Niel de Beaudrap when their initial analysis of the noise model \cite{Beaudrap} showed dephasing to be the major source of error.} \cite{Nielsen2000}. Recall that in the numerical experiments of Section \ref{sec:latice_experiment}, we only used a single qubit from each ion-trap. This means that we could use three qubits from the ion-trap to implement one round of phase-flip code, which would reduce the dephasing noise. By using such a simple phase-flip code we obtained an effective improved dephasing rate of $\approx 2.3\times 10^{-4}$ per second from the one of NQIT noise-level $\approx 7.2\times 10^{-3}$ per second. The results for outputs with far from uniform probabilities are found in \figref{fig:jensFastVsFastNoisyVarious2Deph}.

In this case, roughly half of the runs considering the error corrected dephasing pass our test that the probabilities should be at least within one standard deviation of the perfect run, and greater than one standard deviation of the uniform distribution. This demonstrates partial improvement while being inconclusive as a demonstration of the potential for quantum advantage. In this case an analysis of the \lone{} is particularly valuable.

The readers will find the data required for such an analysis below. In this case, as in the case of the previous figure of merit, a large improvement can be achieved by utilising a simple repetition code. However this improvement might not be as significant as one might expect having seen the results of \figref{fig:jensFastVsFastNoisyVarious2Deph} with the \lone{} still being significantly far from the $\frac{1}{22}$ value required by \theref{2D-DQS hardness}. Indeed even without dephasing noise the \lone{} its too high to expect a demonstration of quantum-advantage. As such we expect both improved error correction codes and error correction applied to other noise channels are required for a demonstration of quantum-advantage.
\begin{center}
    \begin{tabular}{ c|c|c } 
        full noise levels & with repetition code & without dephasing noise \\ 
        \hline
        0.321564704 & 0.270893212 & 0.0656717 \\ 
\end{tabular}
\end{center}

This is a partial improvement relative to the uncorrected results, and so we find a demonstration of quantum-advantage using this error correction scheme as more likely than in the uncorrected case. However, further improvements are required for such a demonstration. 

\subsubsection{The Impact of Continuous Noise Reduction}
\label{sec:TV proxy}

\begin{figure}
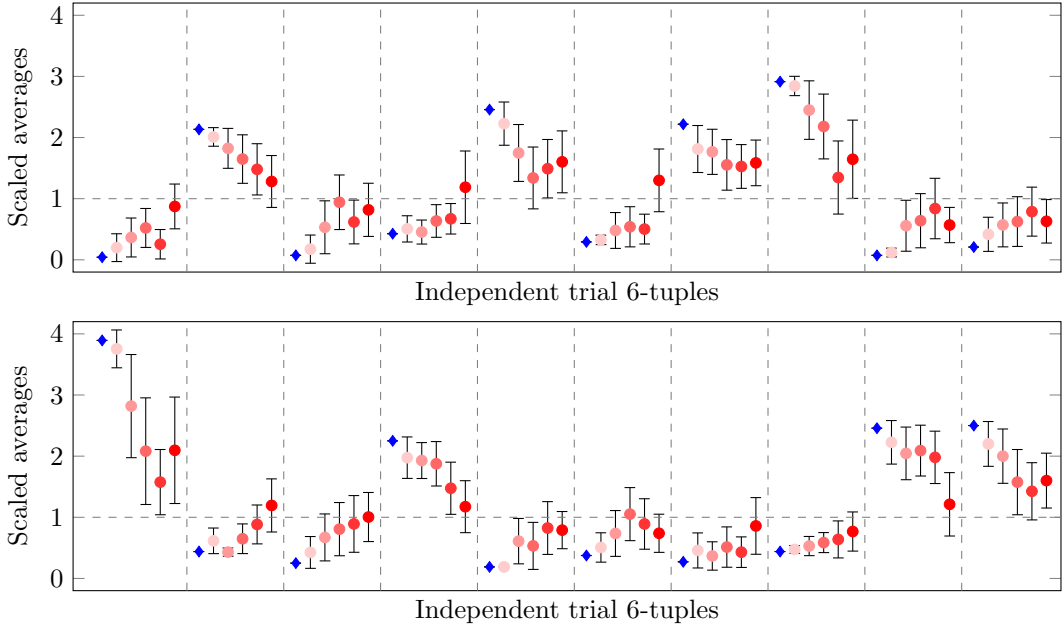

	\centering
	\myinclude{1}{0.3}{jensFastVsFastNoisyVariousDephStages_1}
	\myinclude{1}{0.3}{jensFastVsFastNoisyVariousDephStages_2}
	\caption{Results including reduced dephasing noise rates for a $4\times 5$ ion trap grid. The results referenced by this plot are the probability of measuring a randomly chosen output string, where each trial has a different initial 2D-DQS circuit, and different output string. Every independent trial is described by a 6-tuple, from left to right, of a perfect run (no noise) (blue diamond), the mean of 20 noisy runs with no dephasing errors, the mean of 20 noisy runs with $\frac{1}{4}$ of the NQIT dephasing rate, the mean of 20 noisy runs with $\frac{1}{2}$ of the NQIT dephasing rate, the mean of 20 noisy runs with $\frac{3}{4}$ of the NQIT dephasing rate, the mean of 20 noisy runs with the NQIT dephasing rate. The error bars show one standard deviation. The means and standard deviations have been normalised by the respective uniform distribution (dotted horizontal line).}
	\label{fig:jensVariousNoisyDeph}
\end{figure}

More generally than testing a single error correction code, we can understand how the likelihood of a demonstration of quantum-advantage is affected with a continuously varying noise parameter. Here we will consider dephasing errors, which we have identified as the most damaging form of error. This continuous variation corresponds to, for example, reductions in the gate application time, improvements in the compilation methods or the improved storage of quantum states. The results of this experiment are shown in \figref{fig:jensVariousNoisyDeph}.

While \figref{fig:jensVariousNoisyDeph} appears to demonstrate the continuous improvement which can be achieved by reducing the dephasing error, it seems that it cannot be said that the amplitudes are regularly within one standard deviation of the perfect run until the dephasing rate is reduced to $0$. We do however see that, with regards to our accurate and far from uniform condition, a demonstration of quantum supremacy does become continuously more likely as the dephasing error rate is reduced.

This fact is reinforced by \figref{fig:varyingDephRandCircTV} which shows the average difference between the perfect and noisy runs for each of the values of dephasing error rate. We can use this as a proxy measure for the \lone{}, as discussed in the experimental design methodology introduced in \secref{sec:results}, and as was done earlier in \secref{sec:projections}. In this case we can say that an experiment has a reasonable chance of demonstrating quantum-advantage if we can be convinced that the \lone{} between the noisy and perfect implementations is bounded by $\frac{1}{22}$ which is demanded by the hardness result for the 2D-DQS algorithm as seen in \theref{2D-DQS hardness}. As we do not have access to the full characterisation of the probability distributions, here we will approximate the \lone{} by taking the average difference and proposing that it is representative of the full distribution by scaling it by the uniform distribution. 

We see that even in the case of $0$ dephasing error, the \lone{} is not brought within the $\frac{1}{22}$ value. Instead The average difference in that case is approximately $0.155$ which is significantly higher. However, by our figure of merit, a demonstration of quantum-advantage is made continuously more likely by this fall in dephasing error, showing the advantage in endeavouring to achieve such a fall.

\begin{figure}
	\centering
	\myinclude{0.5}{0.425}{jensFastVsFastNoisyVariousDephStagesVarDistance}
	\caption{Results for varying dephasing noise rates for a $4 \times 5$ ion trap grid. The results referenced by this plot are the difference between the probability amplitudes in noisy and perfect runs when measuring a randomly chosen output string of a random 2D-DQS circuit. The error bars show one standard deviation. The means and standard deviations have been normalised by the uniform distribution.}
	\label{fig:varyingDephRandCircTV}
\end{figure}

\begin{figure}
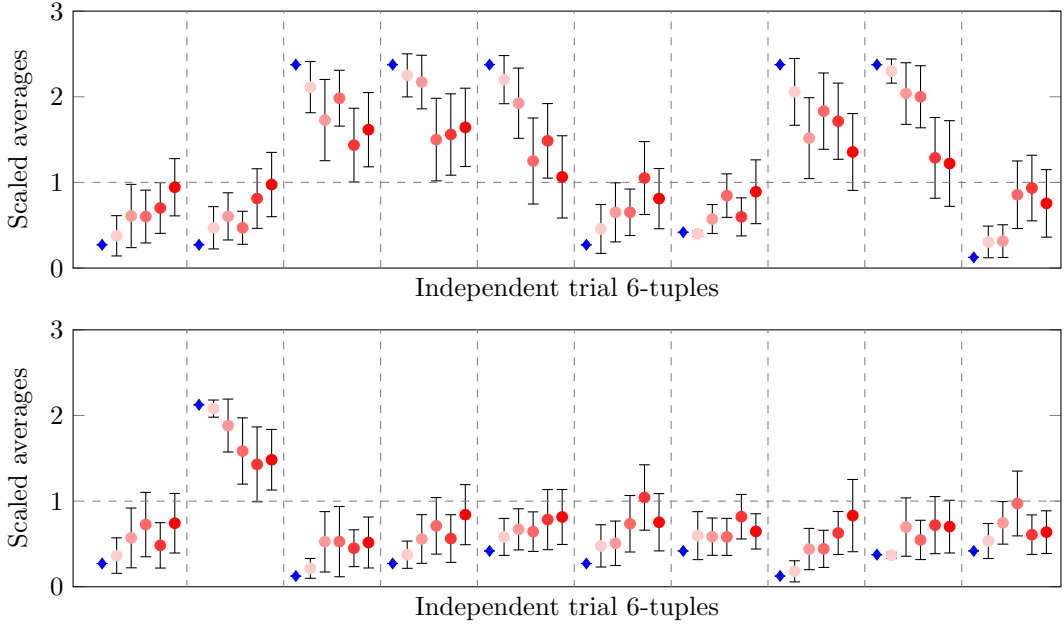

	\centering
	\myinclude{1}{0.3}{jensFastVsFastNoisyVariousDephStagesSameCirc_1}
	\myinclude{1}{0.3}{jensFastVsFastNoisyVariousDephStagesSameCirc_2}
	\caption{Results including reduced dephasing noise rates for a $4 \times 5$ ion trap grid. The results referenced by this plot are the probability of measuring a randomly chosen output string, where each trial has the same initial 2D-DQS circuit, and different output string. Every independent trial is described by a 6-tuple, from left to right, of a perfect run (no noise) (blue diamond), the mean of 20 noisy runs with no dephasing errors, the mean of 20 noisy runs with $\frac{1}{4}$ of the NQIT dephasing rate, the mean of 20 noisy runs with $\frac{1}{2}$ of the NQIT dephasing rate, the mean of 20 noisy runs with $\frac{3}{4}$ of the NQIT dephasing rate, the mean of 20 noisy runs with the NQIT dephasing rate. The error bars show one standard deviation while. The means and standard deviations have been normalised by the respective uniform distribution (dotted horizontal line).}
	\label{fig:varyingDephSameCirc}
\end{figure}

An alternate proxy measure for the \lone{} is to explore the differences between the noisy and perfect amplitudes for a selection of different output bit strings of the same circuit. In \figref{fig:varyingDephSameCirc}, every trial considers the same 2D-DQS circuit, but measures the probability amplitude of a different output bit string.

Once again, this plot can be examined further by directly studying the average difference between the noisy and perfect runs. This proxy measure for the \lone{} is plotted in \figref{fig:varyingDefSameCircTV} but once again the value of $0.135$ is more than twice the $\frac{1}{22}$ which is demanded by the hardness result for the 2D-DQS algorithm as seen in \theref{2D-DQS hardness}.

\begin{figure}
	\centering
	\myinclude{0.5}{0.425}{jensFastVsFastNoisyVariousDephStagesVarDistanceSameCirc}
	\caption{Results for varying dephasing noise rates for a $4 \times 5$ ion trap grid. The results referenced by this plot are the difference between the probability amplitudes in noisy and perfect runs when measuring a randomly chosen output string from a single 2D-DQS circuit. The error bars show one standard deviation. The means and standard deviations have been normalised by the uniform distribution.}
	\label{fig:varyingDefSameCircTV}
\end{figure}

In conclusion, while it seems that reducing dephasing error alone will not be enough to bring a demonstration of quantum-advantage using this scheme within reach, we have seen that utilising a simple 3 qubit correction code would result in a significant improvement on the noise levels. As such we recommend that this error correction technique is used in conjunction with other techniques, correcting for other error types. 

We expect, however, that as the system size grows the \lone{} between the perfect and noisy circuits will grow as the noise modelled is constant for each gate and qubit. This would push a demonstration of quantum-advantage further away, which is consistent with the theoretical results in \cite{Bremner2017achievingsupremacy}. There the authors demonstrate that samples can be efficiently drawn by a classical computer from a distribution produced by an \IQP{} circuits subject to independent depolarising noise on each qubit at the end of the circuit. In that case, however, they show that error correction can be used to recover classical impossibility, if one allows for more complex connectivity, or several rounds of swap gates. While we have restricted the connectivity and circuit depth in our case, there may be gains to be made by removing these restrictions.

\section{Discussion} 
\label{sec:discussion}

We have examined classical simulation of small instances of realistic quantum-advantage computations. The motivation is not to obtain solutions to the problems considered, but to faithfully model the physical system and computation device. 

Having achieved a faithful modelling of the system, classical simulations can be used as a tool in two ways. Firstly, we can use them to benchmark a given device by confirming that the effect of the modelled noise scales correctly. Then, if instances increase in size and continue to match outcomes of real experiments, we extrapolate that the same is true for the, non classically simulatable, quantum-advantage regime. 

The second use is to examine the impact of varying the noise and other constraints and imperfections. By doing so one can identify which limitations contribute most to the degradation of the results, compared to the perfect case. We can then provide feedback to experimentalists as to which aspects of their system they should prioritise in improving, in order to achieve the best results in the specific problem considered. 

We gave a methodology for using classical simulations in the way described above, and exemplified this methodology with two examples, without performing exhaustive explorations of either. In both cases, we considered \IQP{} problems, one of the prominent candidates for demonstrating quantum-advantage. The constraints we imposed were those from the NQIT Q20:20 device \cite{NQIT, Beaudrap}, while the classical simulator used was the one developed by Bravyi and Gosset in \cite{Bravyi2016}. 

The first example used was a subclass of \IQP{} instances, called the 2D-DQS problem and defined in \cite{Bermejo-Vega2017}, with the main focus being the effect of noise. While current NQIT levels of noise are too high, by using our technique we identified that dephasing noise is the most significant source of errors. This led us to a potential solution to improve such computations, namely to use a small phase-flip code to protect from precisely this type of errors, which we showed provided improvements. We also showed that a continuous improvement in the likelihood of a demonstration of quantum-advantage can be achieved by a continuous improvement in the dephasing noise levels. However, we also showed that correcting dephasing error alone would not be sufficient to demonstrate a quantum-advantage using the 2D-DQS protocol on NQIT hardware.

In the second example, we considered a generic IQP-MBQC problem with constraints coming, this time, from architectural limitations. This example was to illustrate how to model different architectures in our framework. We noticed that the current level of noise of NQIT was even more destructive than in the first example.

We give several directions for future research, both specific to the examples considered and more general involving the methodology developed. In Section \ref{sec:results} we provide a tool for benchmarking the Q20:20, but to do such benchmarking, one needs to run these examples on the NQIT Q20:20 and compare with the modelling we obtained. This is naturally the first next step complementing our work. A second direction is to derive theoretical prediction for the effect of noise on our examples, for our problems and with our constraints. This continues the work of Bremner et al \cite{Bremner2017achievingsupremacy} and lets us consider what is required to achieve a demonstration of quantum-advantage.

We should use the methodology developed for using classical simulations in the quantum-advantage problem, in different physical systems and for different problems. For example, it may be beneficial to run though the same benchmarking and prediction process for more general gate and state preparation fidelity estimations \cite{1464-4266-7-10-021, PhysRevLett.106.230501}. Moreover, the use of these simulations as a tool for guiding future experiments should be made more systematic. In Section \ref{sec:projections} we varied the noise starting from coarser grouping of the noise-sources and going to a `\emph{finer-graining}' in order to identifying the major source of errors. We recommend modelling the reduction of a mixture different noise sources as we have shown that removing only one, namely dephasing, would not be sufficient. This could also be enhanced with other techniques, which may also vary the architecture. We envision, that one could use machine-learning techniques to identify, for a given system and problem, the settings that provide the best results with small (to be quantified) improvements.

\paragraph{Acknowledgements:} We are grateful to Niel de Beaudrap for sharing with us their initial work on the effectiveness of small error correcting codes for NQIT device and the details of Q20:20 noise model \cite{Beaudrap}. D.M. would also like to thank the Atos quantum computing R\&D team of Les Clayes-sous-Bois, Paris, for a bountiful supply of expertise in classical simulation of quantum computers. The authors would also like to thank the anonymous reviewers of this work who, though their suggestions, contributed significantly. This work was supported by EPSRC grants EP/N003829/1 and EP/M013243/1, and EP/L01503X/1 for the University of Edinburgh, School of Informatics, Centre for Doctoral Training in Pervasive Parallelism.

\bibliography{bibliography}
\bibliographystyle{unsrtnat}

\appendix

\section{Expanded Circuit Descriptions}

\subsection{IQP-MBQC Circuit in NQIT Gate Set}
\label{sec:IQP_MBQC_circuit}

As discussed in \secref{sec:IQP in MBQC}, for constant $\theta = \pi/8$, each \IQP{} instance is fully defined by a binary matrix $ \mathbf{Q} \in \mathbb{F}_{2}^{n_{g} \times n_{a}}$. For example, $\mathbf{Q}$ of \figref{fig:bipartite graph} corresponds to the circuit
\begin{equation}
    \label{eq:XprogramExample}
    C_X = exp\left(i \frac{\pi}{8} X_1 X_3\right) exp\left(i \frac{\pi}{8} X_2 X_3\right).
\end{equation}

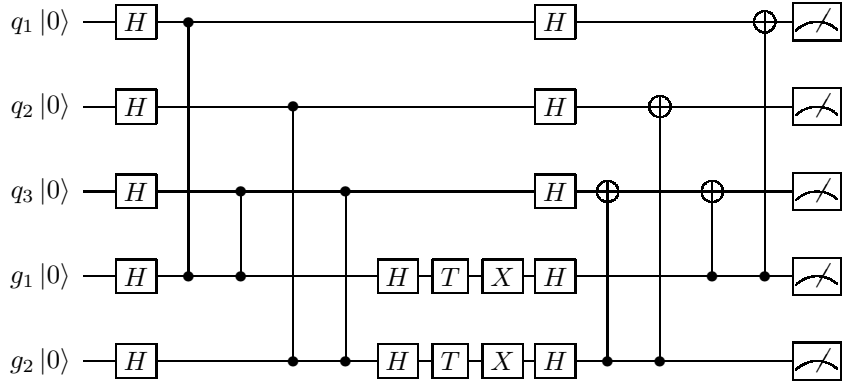
\begin{figure}
	\centering
	$$
	\Qcircuit @C=0.16em @R=1.4em @! {
		\lstick{q_1 \ket{0}} & \gate{H} & \ctrl{3}  & \qw       & \qw       & \qw      & \qw & \qw & \qw                & \gate{H} & \qw       & \qw       & \qw       & \targ     & \meter \\
		\lstick{q_2 \ket{0}} & \gate{H} & \qw       & \qw       & \ctrl{3}  & \qw       & \qw & \qw & \qw                & \gate{H} & \qw       & \targ     & \qw       & \qw       & \meter \\
		\lstick{q_3 \ket{0}} & \gate{H} & \qw       & \ctrl{1}  & \qw       & \ctrl{2}  & \qw & \qw & \qw                & \gate{H} & \targ     & \qw       & \targ     & \qw       & \meter \\
		\lstick{g_1 \ket{0}} & \gate{H} & \ctrl{0}  & \ctrl{0}  & \qw       & \qw       & \gate{H} & \gate{T} & \gate{X} & \gate{H} & \qw       & \qw       & \ctrl{-1} & \ctrl{-3} & \meter \\
		\lstick{g_2 \ket{0}} & \gate{H} & \qw       & \qw       & \ctrl{0}  & \ctrl{0}  & \gate{H} & \gate{T} & \gate{X} & \gate{H} & \ctrl{-2} & \ctrl{-3} & \qw       & \qw       & \meter
	}
	$$
	\caption{Circuit which implements the MBQC pattern of \figref{fig:bipartite graph}. Measurements have been delayed until the end. The final $C X$ gates perform the necessary adaptive corrections.}
	\label{fig:XprogramExampleCircuit}
\end{figure}

To sample from $\bra{0^{\otimes n}} C_X \ket{0^{\otimes n}}$, as is the definition of \IQP{} in \defref{def:IQP oracle}, we must measure the gate qubits in the $\{\ket{0_{\pi/8}}, \ket{1_{\pi/8}}\}$ basis of \eqref{equ:IQP measuremnt basis}. The correct rotation for the $\{\ket{0_{\theta}}, \ket{1_{\theta}}\}$ basis is given by $H R_{-\theta} X R_\theta H$. However, we notice:
\begin{equation}
R_{-\theta} X R_\theta = e^{-i\theta} X R_{2\theta}
\end{equation}
where the global phase can be dropped. Hence the correct rotation for $\theta = \pi/8$ is:
\begin{equation} \label{eq:MBQCgateQubitRotation}
H X R_{2\pi/8} H = H X R_{\pi/4} H = H X T H
\end{equation}

As shown in \cite{Broadbent2009}, we can incorporate the corrections into the circuit by adding $C X$s according to the same pattern used to produce the resource state initially. Since those corrections do not need to be physically executed, because of their equivalence to classical post-processing, we consider them as perfect, i.e. do not add any noise to them. We conclude that the corresponding MBQC pattern, also in \figref{fig:bipartite graph}, can be written in circuit form as in \figref{fig:XprogramExampleCircuit}. This describes an implementation of \IQP \ using the gate set which is available to the NQIT device as discussed in \secref{sec:NQIT_arc} and is the circuit we will implement in our simulator.

\subsection{NQIT Noise Functions}
\label{app:noise functions}

In \algref{noise} we give the implementation of the NQIT noise model of \secref{sec:noise} in the gate based model, which may be understood by the simulator.

\begin{algorithm}
	\caption{Gate based description of the NQIT noise to be used by the simulator. Here $P \left( \lambda \right)$ is a Poisson distribution with mean $\lambda$. The variables listed here assume the current NQIT noise levels but are altered in our experiments of \secref{sec:projections} and can be set to $0$ in the perfect case.}
	\label{alg:noise}
	\begin{algorithmic}[1]
		\State TimeInTrapOperation = $0.5$ms
		\State TimeLinkingOperation = $1.5$s
		\State TimePreparation = $1.25$ms
		\State TimeMeasurement = $2.25$ms
		\State
		\State ProbTwoQubitOperationSingleQubit = $5.5 \times 10^{-5}$
		\State ProbTwoQubitOperationTwoQubit = $6 \times 10^{-5}$
		\State ProbSingleQubitOperation = $1.5 \times 10^{-6}$
		\State ProbMeasurement = $5 \times 10^{-4}$
		\State ProbPreparation = $2 \times 10^{-4}$
		\State ProbDephasing = $7.2 \times 10^{-3}$
		\State ProbDepolarising = $9 \times 10^{-3}$
		
		\breakalg

		\Function{RandomPauli}{$i$,$p$}
			\State Act a Pauli gate, selected uniformly at random, on qubit $i$ with probability $p$
		\EndFunction
		\State
		\Function{DephasingNoise}{$t$, $q$}
			\State Act $Z \left( q \right)$ with probability $P$($t$ProbDephasing)
		\EndFunction
		\State
		\Function{DepolarisingNoise}{$t$, $q$}
			\State \Call{RandomPauli}{$q$, $P$($t$ProbDepolarising)}
		\EndFunction
		\State
		\Function{TimeBaseNoise}{$t$}
			\ForAll{$q \in$ qibits} \Comment{Noise acts on all qubits}
				\State \Call{DephasingNoise}{$t$, $q$}
				\State \Call{DepolarisingNoise}{$t$, $q$}
			\EndFor
		\EndFunction
		\State
		\Function{TwoQubitNoise}{$i,j$}
			\State \Call{RandomPauli}{$i$, ProbTwoQubitOperationSingleQubit}
			\State \Call{RandomPauli}{$j$, ProbTwoQubitOperationSingleQubit}
			\State Act $Z \left( i \right) \otimes Z \left( j \right)$ with probability ProbTwoQubitOperationTwoQubit
		\EndFunction
		\State
		\Function{SingleQubitNoise}{$q$}
			\State \Call{RandomPauli}{$q$, ProbSingleQubitOperation}
		\EndFunction
		\State
		\Function{PreparationNoise}{q}
			\State Act $X \left( q \right)$ with probability ProbPreparation
		\EndFunction
		\State
		\Function{MeasurementNoise}{$q$}
			\State Act $X \left( q \right)$ with probability ProbMeasurement
		\EndFunction
	\end{algorithmic}
\end{algorithm}

Noise is added to a circuit in the following way. All operations are considered independently. Noise gates corresponding to operation based errors are inserted at an operation's position in the circuit at random, with type and probability according to the rates of \secref{sec:noise}. For each of those operations, a nested loop iterates over all qubits in the system and randomly applies the two time-based errors. First the execution time needed for the current operation is calculated by considering the times given in \secref{sec:noise}. Then, at each qubit in the loop, an appropriate noise gate is added according to a Poisson process with the rates listed, again, in \secref{sec:noise}.

\section{Numerical Experiment Details}

\subsection{IQP-MBQC Experiments}
\label{app:IQPMBQC}

\subsubsection{Simulator Benchmarking Experiment of \secref{sec:latice_experiment}}
\label{app:IQPMBQC brute}

Generating random unrestricted \IQP{} instances is equivalent to randomly populating $\mathbf{Q}$ with zeros and ones. The description in \apref{sec:IQP_MBQC_circuit} of how to convert a given X-program $\mathbf{Q}$ to a particular circuit lets us control the T-gates count $t$. We saw that every individual exponential (row in $\mathbf{Q}$) corresponds exactly to $t=1$, and the number of application qubits has no effect on $t$. We want T-gate counts of no more than $20$ in order to achieve feasible run-times.

One trial consists of generating a random \IQP{} instance, obtaining the true probability of measuring the $\ket{0}^{n}$ using brute-force, and solving them with the simulator of \cite{Bravyi2016} 20 times. Each instance is created by randomly populating with binary values a matrix $\mathbf{Q}$ of randomly picked dimensions in $[5,15]\times [5,12]$. This corresponds to $n \in [5,12]$ and $t \in [5,15]$ where the complexity in the brute-force case is determined by $n$, and in the simulator's, by $t$.

The experiment consists of $20$ trials, with the mean of the simulator output in each trial compared to the brute force case to give the coefficient of determination.

\subsubsection{NQIT Noise and Architecture Restricted Experiment of \secref{sec:NQIT_experiment}}
\label{app:IQPMBQC NQIT}

We again generate random IQP-MBQC circuits, but under the restrictions described in \secref{sec:NQIT_experiment}. Rather than a full matrix, $\mathbf{Q}$, it is now sufficient for each gate qubit, $g_i$, in an ion trap, $i$, to have corresponding bit strings, $i^0$ and $i^1$, indicating the entanglement patterns between itself and qubits in it and its neighbouring ion trap.

Details of the circuit simulated can be seen in \algref{noisy MBQC}. Once the circuit is simulated, we calculate the probability that an NQIT implementation would measure a random bit string $b$. One noisy run consists of simulating the circuit produced from \algref{noisy MBQC}, using fixed $i^0,i^1,b$, $20$ times to calculate the mean and standard deviation. Then a new tuple $i^0, i^1, b$ is generated and the process is repeated for the next trial. A perfect run is equivalent but with the noise values set to $0$, with the perfect and noisy pair forming one trial. In total the experiment consists of $20$ trials.

\begin{algorithm}
	\caption{Code producing a noisy IQP-MBQC circuit, to be implemented by the simulator, as discussed in \secref{sec:NQIT_experiment}. We use $i$ to index the ion traps, and to represent the set of $K' - 2$ available application qubits which each trap contains ($K'$ minus 1 qubit $c_i$ to receive the gate qubit from it's neighbour, minus one gate qubit $g_i$).}
	\label{alg:noisy MBQC}
	\begin{algorithmic}[1]
		\Require For every ion trap, $i$, two strings, $i^0$, $i^1$. Bit string $b$.
		\Ensure Noisy circuit.
		\ForAll{$q \in$ qubits}
			\State \Call{Initialise}{$q$} \Comment{Recall, initialisation is in the $\ket{+}$ state}
			\State \Call{PreparationNoise}{$q$}
		\EndFor
		\State
		\ForAll{$i \in$ ion traps, except the last}
			\ForAll{$q \in i$}
				\If{$i_{q}^{0} = 1$}
					\State Act $CZ \left( g_i , q \right)$
					\State \Call{TwoQubitNoise}{$g_i$, $q$}
					\State \Call{TimeBasedNoise}{TimeInTrapOperation}
				\EndIf
			\EndFor
		\EndFor
		\State
		\ForAll{$i \in$ ion traps, except the last, such that $i$ is even}
			\State \Call{Swap}{$g_i$, $c_{i+1}$} \Comment{Move gate qubits to neighbouring ion trap}
		\EndFor
		\State \Call{TimeBasedNoise}{TimeLinkingOperation + TimeMeasurement}
		\State
		\ForAll{$i \in$ ion traps, except the last, such that $i$ is odd}
			\State \Call{Swap}{$g_i$, $c_{i+1}$}
		\EndFor
		\State \Call{TimeBasedNoise}{TimeLinkingOperation + TimeMeasurement}
		\State
		\ForAll{$i \in$ ion traps, except the first}
			\ForAll{$q \in i$}
				\If{$\left(i-1 \right)_{q}^{1} = 1$}
					\State Act $CZ \left( g_{i-1} , q \right)$
					\State \Call{TwoQubitNoise}{$g_{i-1}$, $q$}
					\State \Call{TimeBasedNoise}{TimeInTrapOperation}
				\EndIf
			\EndFor
		\EndFor
		\breakalg
		\ForAll{$i \in$ ion traps, except the first}
			\State Act $H \left( g_{i-1} \right)$
			\State \Call{SingleQubitNoise}{$g_{i-1}$}
		\EndFor
		\State \Call{TimeBasedNoise}{TimeInTrapOperation}
		\ForAll{$i \in$ ion traps, except the first}
			\State Act $T \left( g_{i-1} \right)$
			\State \Call{SingleQubitNoise}{$g_{i-1}$}
		\EndFor
		\State \Call{TimeBasedNoise}{TimeInTrapOperation}
		\ForAll{$i \in$ ion traps, except the first}
			\State Act $X \left( g_{i-1} \right)$
			\State \Call{SingleQubitNoise}{$g_{i-1}$}
		\EndFor
		\State \Call{TimeBasedNoise}{TimeInTrapOperation}
		\State
		\ForAll{$q \in$ qubits}
			\State Act $H \left( q \right)$
			\State \Call{SingleQubitNoise}{$q$}
		\EndFor
		\State \Call{TimeBasedNoise}{TimeInTrapOperation}
		\State
		\ForAll{$i \in$ ion traps, except the last} \Comment{CNOT seen at end of \figref{fig:XprogramExampleCircuit}}
			\ForAll{$q \in i$}
				\If{$i_{q}^{0} = 1$}
					\State Act $CX \left( g_{i} , q \right)$
				\EndIf
			\EndFor
		\EndFor
		\State
		\ForAll{$i \in$ ion traps, except the first}
			\ForAll{$q \in i$}
				\If{$\left( i - 1 \right)_{q}^{1} = 1$}
					\State Act $CX \left( g_{i-1} , q \right)$
				\EndIf
			\EndFor
		\EndFor
		\State
		\ForAll{$i \in$ ion traps}
			\ForAll{$q \in i$ and $g_{i-1}$ for all but the first ion trap}	
				\State \Call{MeasurementNoise}{$q$}
			\EndFor
		\EndFor
		\State \Call{Measure}{$b$} \Comment{Give the probability of measuring $b$ in the Computational basis}
	\end{algorithmic}
\end{algorithm}

Notice that we are being pessimistic in \algref{noisy MBQC} by assuming that there is no parallelism in the gate applications. As such we apply time based noise after each gate. We have also simplified the operation of swapping to a single operation, rather than a protocol as seen in \algref{entangleTwoTraps}. This reduces the simulation time while roughly maintaining the noise impact, as the time based noise should dominate here.

\subsection{2D-DQS Experiments of \secref{sec:NQIT noise experiment} and \secref{sec:projections}}
\label{sec:appendixJens}

When entangling the traps to form the resource state, we extend the procedure shown in \eqref{eq:trapsTwoSteps}. Instead of only 2-steps, as it is in the 1D case, we need 4 steps for a 2D grid resource state. We achieve this by entangling sequentially:
\begin{itemize}
	\item Even-indexed columns' qubits to their right neighbours
	\item Odd-indexed columns' qubits to their right neighbours
	\item Even-indexed rows' qubits to their bottom neighbours
	\item Odd-indexed rows' qubits to their bottom neighbours
\end{itemize}

Having performed the entanglement we are left to apply the $T$-gates and measure. We track the qubits on which we apply the $T$ gates using the bit string $\tau$ which takes the value $1$ at the locations where a $T$ gate is applied.

We calculate the amplitude of a randomly selected output, $b$, for each instance in order to simulate sampling. We calculate several trials where for each we:
\begin{itemize}
	\item Generate a uniformly random $\tau \in \left[ 0 , 1 \right]^{20}$ to give a 4x5 circuit as in \secref{sec:2d-dqs}. 
	\item Generate a random bit string, $b$, to calculate the amplitude of.
	\item Solve $20$ times and take the mean and standard deviation. This is a perfect run.
	\item Generate 20 random noisy circuits, one per noisy run, based on the perfect one by inputting $\tau$ into \algref{noisy NQIT}. In the case of \secref{sec:projections} we will use different values for the variables of \algref{noise}, as discussed there.
	\item For each noisy run, solve the circuit $20$ times and calculate the mean. The result is a vector of length $20$ containing these mean values.
\end{itemize}

Attempts to reduce the standard deviation of the noisy runs by increasing the number of times the computation is performed during each run were not effective, suggesting the deviation is a result of the noise.

\begin{algorithm}
	\caption{Code producing a noisy 2D-DQS circuit, to be implemented by the simulator, as discussed in \secref{sec:NQIT noise experiment} and \secref{sec:projections}. We will index traps (and equivalently, in this case, qubits) by the row, $n$, and column, $m$, where then appear in the square grid. }	
	\label{alg:noisy NQIT}
	\begin{algorithmic}[1]
		\Require Bit strings $\tau$ and $b$.
		\Ensure Noisy circuit.
		\ForAll{$q \in$ qubits} \Comment{This and the following loop initialise $\ket{+}$ states}
			\State \Call{Initialise}{$q$}
			\State \Call{PreparationNoise}{$q$}
		\EndFor
		\State
		\For{$p \in \left\{ odd , even \right\}$} \Comment{Entangle columns of lattice}
			\For{$\left\{n,m : n \in p \right\}$}
				\State $CZ \left( \left( n , m \right) , \left( n + 1 , m \right) \right)$
				\State \Call{TwoQubitNoise}{$\left( n , m \right) , \left( n + 1 , m \right)$}
				\State \Call{TimeBasedNoise}{TimeInTrapOperation}
			\EndFor
		\EndFor
		\State
		\For{$p \in \left\{ odd , even \right\}$} \Comment{Entangle rows of lattice}
			\For{$\left\{n,m : m \in p \right\}$}
				\State $CZ \left( \left( n , m \right) , \left( n , m + 1 \right) \right)$
				\State \Call{TwoQubitNoise}{$\left( n , m \right) , \left( n , m + 1 \right)$}
				\State \Call{TimeBasedNoise}{TimeInTrapOperation}
			\EndFor
		\EndFor
		\State
		\For{$q \in qubits$} \Comment{Act $T$ gate according to original circuit}
			\If{$\tau_{i} = 1$}
				\State Act $T \left( i \right)$
				\State \Call{SingleQubitNoise}{$i$}
				\State \Call{TimeBasedNoise}{TimeInTrapOperation}
			\EndIf
		\EndFor
		\State
		\For{$q \in qubits$}
		    \State Act $H \left( q \right)$ \Comment{Ajust to measure in the Hadamard basis}
		    \State \Call{SingleQubitNoise}{$q$}
			\State \Call{TimeBasedNoise}{TimeInTrapOperation}
		\EndFor
		\State
		\For{$q \in qubits$}
			\State \Call{MeasurementNoise}{$q$}
		\EndFor
		\State \Call{Measure}{$b$}
	\end{algorithmic}
\end{algorithm}

\end{document}